\documentclass[11pt]{article}

\usepackage{setspace}
\usepackage[utf8]{inputenc}
\usepackage[english]{babel}
\usepackage{csquotes}
\usepackage{authblk} 
\usepackage{amsmath, amsthm, amssymb}

\usepackage{acro}
\usepackage[inline,shortlabels]{enumitem}
\usepackage{caption}
\usepackage{subcaption}
\usepackage{booktabs}      
\usepackage{tabularx}
\usepackage{pdflscape}
\usepackage{siunitx}
\usepackage{dcolumn}
\usepackage[most]{tcolorbox}
\usepackage{xcolor}
\usepackage{longtable}
\usepackage{xltabular}
\usepackage{multicol}
\usepackage{multirow}
\usepackage{makecell}

\usepackage{geometry}
\usepackage{listings} 
\usepackage{color}
\definecolor{lightgray}{rgb}{.9,.9,.9}
\definecolor{darkgray}{rgb}{.4,.4,.4}
\definecolor{purple}{rgb}{0.65, 0.12, 0.82}

\lstdefinelanguage{JavaScript}{
  keywords={typeof, new, true, false, catch, function, return, null, catch, switch, var, if, in, while, do, else, case, break, event},
  keywordstyle=\color{blue}\bfseries,
  ndkeywords={class, export, boolean, throw, implements, import, this},
  ndkeywordstyle=\color{darkgray}\bfseries,
  identifierstyle=\color{black},
  sensitive=false,
  comment=[l]{//},
  morecomment=[s]{/*}{*/},
  commentstyle=\color{purple}\ttfamily,
  stringstyle=\color{red}\ttfamily,
  morestring=[b]',
  morestring=[b]"
}

\usepackage[round]{natbib}

\usepackage{hyperref}
\hypersetup{colorlinks=true,
	linkcolor=blue,
	citecolor=blue,
	urlcolor=blue}

\addto\extrasenglish{%
}	
\addto\extrasenglish{%
}

\usepackage{algorithm}
\usepackage{algorithmic}
	
\usepackage{tabularx}
\usepackage{booktabs}
\usepackage{multicol}
\usepackage{graphicx}

\DeclareAcronym{llm}{
    short = LLM,
    long = large language model
}

\DeclareAcronym{afa}{
    short = AFA,
    long = American Finance Association
}

\DeclareAcronym{nber}{
    short = NBER,
    long = National Bureau of Economic Research
}

\DeclareAcronym{wfa}{
    short = WFA,
    long = Western Finance Association
}

\DeclareAcronym{efa}{
    short = EFA,
    long = European Finance Association
}

\DeclareAcronym{cicf}{
    short = CICF,
    long = China International Conference in Finance
}

\DeclareAcronym{api}{
    short = API,
    long = application programming interface
}

\DeclareAcronym{ocr}{
    short = OCR,
    long = optical character recognition
}

\DeclareAcronym{vif}{
    short = VIF,
    long = variance inflation factor
}

\DeclareAcronym{doi}{
    short = DOI,
    long = digital object identifier
}

\DeclareAcronym{jf}{
    short = JF,
    long = Journal of Finance
}

\DeclareAcronym{jfe}{
    short = JFE,
    long = Journal of Financial Economics
}

\DeclareAcronym{rfs}{
    short = RFS,
    long = Review of Financial Studies
}

\DeclareAcronym{assa}{
    short = ASSA,
    long = Allied Social Science Associations
}

\title{How Well Do LLMs Perform in Peer Review?\\ Evidence from Finance and Accounting}

\author{%
Hai Lu\thanks{University of Toronto and Peking University; email: {\tt hai.lu@rotman.utoronto.ca}}%
\hspace{2.5em}%
Yichen Luo\thanks{University College London; email: {\tt yichen.luo.22@ucl.ac.uk}}%
\hspace{2.5em}%
Luping Yu\thanks{University of Macau; email: {\tt lupingyu@um.edu.mo}}%
}

\begin{document}

\maketitle

\begin{abstract}
Referees increasingly use large language models (LLMs), while finance and accounting journals disagree about whether to allow it. We ask which part of a referee's job the models can do. GPT-4o, whose training data end before the 2024 conference season, referees the papers presented at the 2024 AFA and NBER finance meetings, and we compare its verdicts with publication in the top-three finance journals over the next two years. Papers the model scores one standard deviation higher are 44\% more likely to be published, net of author reputation. The result is not driven by look-ahead bias, as the model does not recall the outcomes. The information comes from its judgements of correctness and exposition. Its judgements of originality and significance carry none, and a referee guide that stresses importance does not help. The split between correctness and contribution holds for accounting papers and for a newer model, GPT-5. LLMs can check whether research is right, but not whether it matters. The technical check can go to the model, while the judgement of contribution stays with people.


\end{abstract}
\newpage

\section{Introduction}
\label{sec:introduction}

Generative AI has reached both sides of peer review. It lowers the cost of producing manuscripts \citep{novymarx2026}. At \emph{Organization Science}, submissions rose 42\% after the release of ChatGPT, and the editors attribute nearly all of the increase to heavily AI-written manuscripts \citep{gartenberg2026}. In economics, Paul Goldsmith-Pinkham finds that the share of papers with at least one passage that a detector classifies as AI-written rises from almost zero in 2022 to roughly a quarter of NBER working papers and nearly half of arXiv economics papers, with the acceleration beginning in 2025.\footnote{See \url{https://paulgp.substack.com/p/how-do-we-know-if-writing-is-ai}.} Referees use the same tools, and they are stretched thin. At ICLR 2026, an AI detector classified about 21\% of the reviews as fully AI-generated and found signs of AI use in more than half \citep{naddaf2025iclr}.\footnote{The International Conference on Learning Representations (ICLR) is a leading machine learning conference. Evaluating its 2026 submissions took 18,054 reviewers and 76,139 reviews, about five papers per reviewer in two weeks \citep{naddaf2025iclr}.} More than 30\% of the referee reports at \emph{Organization Science} show some AI use, and the editors warn that AI, combined with the incentive to publish, pushes science toward more rather than better research \citep{gartenberg2026}. If AI multiplies the number of papers and already writes many of the reviews, can it also help tell which papers are better?

Computer science has started to answer this question, because some of its leading conferences publish their reviews. Comments written by GPT-4 overlap with those of human reviewers about as much as two human reviewers overlap with each other \citep{liang2024useful}. The Stanford Agentic Reviewer, calibrated on reviews from ICLR 2025, produces scores that agree with a human reviewer about as well as two human reviewers agree with each other \citep{paperreview2025}. The conferences disagree on whether reviewers may use these models at all, from a ban at CVPR to deployments at AAAI-26 and NeurIPS 2026 (\autoref{tab:llm_policies}). At AAAI-26, authors and committee members rated the \ac{llm} reviews above human reviews at identifying technical errors \citep{biswas2026aaai}.

Business research has no comparable evidence, although its journals are already setting rules, and the rules disagree within each field. In finance, the \emph{Journal of Finance} lets referees use AI to draft or edit their reports and to look up references, provided they fact-check the output and remain accountable for it.\footnote{See \url{https://afajof.org/submissions/}.} Elsevier, which publishes the \emph{Journal of Financial Economics}, bars reviewers from uploading any part of a manuscript to an AI tool and allows the tools only to improve the language of a report or to search the literature, with disclosure.\footnote{See \url{https://www.elsevier.com/about/policies-and-standards/generative-ai-policies-for-journals}.} In accounting, \emph{The Accounting Review} lets reviewers use AI to improve the language of their report, with disclosure, but not to read the manuscript.\footnote{See \url{https://aaahq.org/Research/Journals/The-Accounting-Review/Guide-for-Authors}.} \emph{Contemporary Accounting Research} prohibits reviewers from using the tools at all.\footnote{See \url{https://www.caaa.ca/car-editorial-policies}.} These policies have been continually updated to keep pace with advances in AI technology. An experiment in which an AI agent wrote referee reports on three of the author's own finance papers in under two minutes each has prompted a call for guardrails on referees' and editors' use of these tools \citep{subrahmanyam2026review}. Referee reports at finance and accounting journals are confidential, so no one outside an editorial office can say whether the verdict of an \acs{llm} on a finance paper tracks the judgement of the field, or which part of its verdict does.

We build a proxy benchmark from the one part of the review process that is public: the decision. We take the 363 papers from the 2024 \acs{afa} Annual Meeting and the six \acs{nber} finance program meetings of 2024 whose PDFs the conferences hosted; the 280 outside the \acs{afa} PhD student poster session form the main sample. We ask GPT-4o, whose training data end three months before the \acs{afa} meeting, to referee the conference version. We call our prompt the naive prompt: it is the three sentences a referee who has done no prompt engineering might write, asking for a report, a recommendation, scores for originality, significance, correctness, and exposition, and an overall score. We then observe which papers the \emph{Journal of Finance}, the \emph{Journal of Financial Economics}, and the \emph{Review of Financial Studies} published within two years of the meeting. The model's scores are ex ante judgements, and the question is whether they carry information about the ex post decisions of editors and referees. Two further tests follow: To ask whether the result is specific to finance, we repeat the design on the 41 accounting papers of the 2023 and 2024 Conferences on Financial Economics and Accounting, a joint accounting and finance conference that hosts the full text of its papers. We use only its accounting papers, not those of its finance sessions. To ask whether a newer model can judge what an older one cannot, we repeat it with GPT-5.

Our key message is that an \acs{llm} referee can tell whether a finance paper is right and clearly written, but not whether it matters. Five findings support it.

First, the model's verdict predicts publication. A one-standard-deviation rise in the model's score makes a paper \emph{44\% more likely} to appear in a top-three finance journal, or 2.9 percentage points from a base rate of 6.5\%. The estimate is net of the authors' school ranking, publication record, and editorial experience. Nor does the model remember the outcomes: when we ask whether each paper has been published, it says it does not know for 90\% of them even when given the authors, and the score keeps its predictive power once its answers are controlled for.

Second, this information lies in the model's judgements of \emph{correctness}, whether the analysis, empirical design, and interpretation are convincing, and of \emph{exposition}, whether the paper is clearly written. Its originality and significance scores carry none, and for top-three publication its overall score predicts less well than its correctness or exposition score alone because it mixes in judgements that carry no information. This is the opposite of what the profession says about its own decisions. In the guide by \citet{berk2017}, most submitted papers are correct and the decision often turns on importance, so the model is informative about the margin that most submissions clear and uninformative about the margin on which the decision turns.

Third, handing the model a referee guide does not help. Given the article by \citet{berk2017} to read before the paper, the model's originality and significance scores still carry no information, the information again lies in its correctness score, and its overall score no longer predicts publication. Telling the model that importance is what matters does not make its verdict more informative.

Fourth, the pattern is not specific to finance. On the 41 accounting papers, the model's overall and correctness scores predict publication in the top-three accounting journals, \emph{The Accounting Review}, the \emph{Journal of Accounting Research}, and the \emph{Journal of Accounting and Economics}, within two years, its exposition score does so at the 10\% level, and its originality and significance scores do not.

Fifth, the limit is not specific to an older model. GPT-5, released in August 2025, shows the same split: its correctness score predicts publication, and its originality and significance scores do not. Its later knowledge cutoff opens the possibility of look-ahead bias, so we repeat the look-ahead probe with GPT-5 and drop the two papers whose journal it names and that were submitted or online before its cutoff. The problem we document is therefore one that a newer model inherits rather than solves.

What can finance and accounting researchers take from these results? We organize the answer around three questions.

The first question is whether a model can replace a human referee. The answer is no, and the reason is more specific than a general distrust of the models. On importance, the margin that editors describe as decisive \citep{berk2017}, the model's judgement carries no information about the decision. Human referees find this margin hard as well: their recommendations on the same paper disagree widely \citep{welch2014}, and the classic papers they once rejected show how poorly importance is judged at the time \citep{gans1994}. But it is the margin the field cannot delegate. The limit is not in the model's reading, since the same reading yields an informative judgement of correctness, and more knowledge does not close it, since a newer model that has read more of the literature shows the same limit. Whether a question matters is a judgement call about what a field run by people will value, and the models do not yet make that call as editors do. Computer scientists reach the same conclusion: an \acs{llm} finds about half of the errors planted in short papers but errs more often than not when asked which of two abstracts is better \citep{liu2023reviewergpt}, and at ICLR 2026 its scores predict acceptance less well than human scores do \citep{baumann2026stop}.

The second question is whether a model can check a paper before a person reads it. The answer is yes, on the margins of correctness and exposition. A referee drafting a report can ask the model whether the design, the evidence, and the writing hold up, and treat the answer as a first reading that a person then verifies. An editor facing a growing volume of submissions can use the same judgement to decide where a referee's time is best spent. An author can run the check before submitting. The division resembles the one found in other knowledge work, where AI helps on tasks inside its competence and hurts outside it \citep{dellacqua2026jagged}, and in security analysis, where a machine's reading of hard data and an analyst's reading of soft information do best combined \citep{cao2024manmachine}. It also speaks to the field's worry that fragile and irreproducible results pass human review \citep{harvey2017, hail2020}. On that part of the job a model is a cheap second reader that does not tire, and work on replication packages shows how much such a reader can catch \citep{schwartz2026}.

The third question is what the results imply for the rules that journals and conferences are writing. The useful question is not whether to admit the model, on which the rules in \autoref{tab:llm_policies} differ, but what to ask of it: the technical check can go to the model, and the judgement of contribution should stay with people. \autoref{sec:implications} develops this for the venues in the table.

We contribute to three literatures. The first studies peer review and publication in economics, finance, and accounting: what editors and referees reward \citep{laband1990, ellison2002, card2020, berk2017}, how much referees agree \citep{welch2014}, how authors' standing and connections and the circulation of a paper before submission shape outcomes \citep{laband1994, colussi2018, brogaard2014, huber2022, brown2005}, and what predicts the publication of finance working papers \citep{dai2023, hou2025}. Peer review and publication are integral to the process of intellectual exploration, shaping not only which research findings enter the scholarly record but also how ideas are evaluated, refined, and advanced. In accounting, \citet{moizer2009} and \citet{bloomfield2018} examine the review process itself, and the editors of \emph{Accounting Horizons} have asked what role people should keep as AI enters academic research \citep{graham2026ai}. We add a measure of content produced by a model that reads the paper, and show which of its judgements align with editors' decisions.

The second literature evaluates \acsp{llm} as referees. In computer science, reviews written by \acsp{llm} are compared with the public reviews of human referees \citep{liang2024useful, paperreview2025, baumann2026stop}, a randomized trial at ICLR 2025 has tested \acs{llm} feedback to referees \citep{thakkar2026randomized}, and AAAI-26 has reported on its pilot of \acs{llm} reviews \citep{biswas2026aaai}. In economics, \citet{pataranutaporn2026} show that \acs{llm} ratings of published papers track the prestige of the journals that published them, and that some models rate papers higher when they are attributed to prominent authors and institutions. \citet{li2026merit} uses a model's reading of working papers to separate the roles of merit and connections in journal placement. In finance and accounting, \citet{newton2026} show that a rubric-guided model scores \acs{afa} papers with high reliability, \citet{novymarx2026} shows that the models can produce finance papers at scale, and \citet{ciconte2026} that AI-drafted accounting papers are polished but not ready for a top journal. We relate the scores of a model given the naive prompt to editorial outcomes and trace the information to its correctness and exposition scores.

The third literature uses \acsp{llm} as tools in finance and accounting research, from forecasting returns and corporate policies \citep{lopezlira2026, chen2026expected, jha2024chatgpt} to reading disclosures and measuring constructs in text \citep{dekok2025}. It studies the effects of the models on information processing in markets \citep{bertomeu2025}, with surveys by \citet{eisfeldt2025generative} and \citet{cao2026genai}, and it has developed the tests for look-ahead bias that we adopt \citep{glasserman2024, lopezlira2025, sarkar2024lookahead, gao2025} and documented the bias in finance and accounting \citep{levy2026caution}. Our results add a task, the evaluation of research, on which the division of labour between model and person can be measured.

\autoref{sec:data} describes the data and design, \autoref{sec:results} reports the results, \autoref{sec:implications} draws out the implications for the rules on AI in peer review, and \autoref{sec:conclusion} concludes. To keep the paper concise and increase its readability, we move the detailed discussions of the sample, the referee protocol, the tracking of publication outcomes, and additional tests to the Internet Appendix.

\section{Data and Design}
\label{sec:data}

The design has three parts: a set of papers whose versions are fixed in time, an \acs{llm} referee that could not have seen what happened to them, and the editorial decisions that followed. It is the validation against realized outcomes, with an explicit test for leakage, that \citet{ludwig2026llm} recommend before an \acs{llm}'s output is used in empirical work. The Internet Appendix gives the details, and \autoref{tab:var_des} defines all variables.

\subsection{Papers}

The main sample consists of every paper on the programs of the 2024 \acs{afa} Annual Meeting (January 2024) and the six 2024 \acs{nber} finance program meetings for which the conference hosts the PDF: 363 papers after duplicates and non-paper files are removed. The \ac{wfa}, \ac{efa}, and \ac{cicf} do not post the full text of their 2024 papers. The main analysis uses the 280 papers outside the \acs{afa} PhD student poster session, a separate track for doctoral students. We use conference programs rather than a working-paper repository because a program timestamps a paper: the file on the program is the version that existed at the meeting, whereas a repository such as SSRN serves only the latest version \citep{dai2023}. The model reads the full conference version, including tables, references, and appendices. Internet Appendix~\ref{sec:appendix_sample} describes the construction of the sample and of the text.

As a robustness check with evidence from another discipline, we apply the same design to the 41 accounting papers presented at the 2023 and 2024 Conferences on Financial Economics and Accounting (Internet Appendix~\ref{sec:appendix_acct}), a joint accounting and finance conference that also hosts the full text of its papers. We use only its accounting papers, not those of its finance sessions. The accounting sample is small because it comes from a single conference: most accounting conferences, the high-quality Financial Accounting and Reporting Section (FARS) Midyear Meeting of the American Accounting Association among them, either do not post the PDFs of the papers they present or restrict them to participants, so the version presented at the meeting cannot be obtained.

\subsection{The LLM Referee}
\label{sec:referee}

The referee is \texttt{gpt-4o-2024-11-20}, whose stated knowledge cutoff is 1 October 2023, three months before the \acs{afa} meeting. Each paper is evaluated in a fresh session, with no other paper in context.

The prompt, which we call the naive prompt, imitates a referee who does not engage in prompt engineering. In three sentences it asks the model to review the paper as a referee for a finance journal, to write a report and recommend acceptance, revision, or rejection, and to score the paper from 0 to 10 on four dimensions and from 0 to 100 overall. The four dimensions, their one-sentence definitions, and the two scales follow \citet{newton2026}:
\begin{itemize}[nosep]
    \item \emph{Originality}: does the paper present a genuinely new idea, angle, dataset, design, or synthesis relative to established academic knowledge?
    \item \emph{Significance}: how important is the question, and how large is the contribution if the results are correct?
    \item \emph{Correctness}: are the analysis, empirical design, theoretical logic, and interpretation scientifically convincing?
    \item \emph{Exposition}: how clear, well organized, and well communicated is the paper?
\end{itemize}
The prompt contains no rubric, no example, and no instruction on how to weigh the dimensions (Internet Appendix~\ref{sec:prompt} reproduces it). For the accounting papers it is addressed to an accounting journal. Because a chat interface gives a different answer each time, we evaluate each paper four times in independent sessions and use the mean of the four draws, standardized across papers, so that a regression coefficient measures the effect of one standard deviation and does not depend on the units of the scale.

\subsection{Publication Outcomes}

The outcome is the editorial decision of the top-three journals of the field. ${\it Top3Fin}_i$ equals one if finance paper $i$ is online at the \emph{Journal of Finance}, the \emph{Journal of Financial Economics}, or the \emph{Review of Financial Studies} within two years of its meeting, the main outcome of \citet{dai2023}. ${\it Top3Acc}_i$ equals one if accounting paper $i$ is online at \emph{The Accounting Review}, the \emph{Journal of Accounting Research}, or the \emph{Journal of Accounting and Economics} within two years of its meeting. A fixed window gives every paper the same time to be accepted, and a paper published in any other journal has an outcome of zero. We match papers to journal articles by title and authors and check every uncertain match by hand, following \citet{hou2025}, so that papers whose titles changed before publication are found. The window had closed for every paper when we collected the outcomes. Of the 276 finance papers of the estimation sample, 18 (6.5\%) are in a top-three finance journal; of the 41 accounting papers, 5 (12.2\%) are in a top-three accounting journal. Internet Appendix~\ref{sec:appendix_outcomes} describes the matching and reports the broader journal sets of \citet{dai2023}.

\subsection{Specification}
\label{sec:spec}

Following Equation~(1) of \citet{dai2023}, we estimate
\begin{equation}
    \Pr\big({\it Top3Fin}_{i} = 1\big)
        = \Lambda\!\left(\alpha \cdot {\it Score}_{i}
        + \boldsymbol{\beta}' \mathbf{X}_{i}
        + \mu_{c}\right),
    \label{eq:logit}
\end{equation}
where $\Lambda$ is the logistic function, ${\it Score}_{i}$ is one of the standardized \acs{llm} scores, and $\mu_{c}$ distinguishes \acs{afa} from \acs{nber} papers. The controls $\mathbf{X}_{i}$ are the log length of the paper, the number of authors, and three measures of the authors' standing: whether any author is at one of the top-twenty finance schools, the prior \acs{jf}, \acs{jfe}, and \acs{rfs} record of the most published author, from OpenAlex (Internet Appendix~\ref{sec:appendix_reputation}), and ${\it Editorial}_i$, one if any author served as an editor or associate editor of a top-three journal of the field in the ten years up to the conference, 2014--2024 \citep{brogaard2014}. For the accounting papers the outcome is ${\it Top3Acc}_i$, the record and editorial variables refer to the three accounting journals, the school indicator is the same as for the finance papers, and the fixed effect distinguishes the 2023 and 2024 meetings. We report robust $z$-statistics; the text reports the average marginal effect of the overall score. The estimation sample of the main specification has 276 papers: of the 280, we drop the paper none of whose authors could be matched to OpenAlex, the paper for which the model names the journal of an article submitted before its cutoff (Internet Appendix~\ref{sec:appendix_lookahead}), and the two papers whose journal GPT-5 names and that were submitted or online before its cutoff (\autoref{sec:results_models}).

\autoref{tab:sum_scores} reports summary statistics. Two features stand out. The model assigns relatively high and uniform overall scores, averaging 81.9 for the finance papers (Panel A), with a standard deviation of only 2.4. In contrast, the dimension scores exhibit greater relative dispersion, as reflected in their higher coefficients of variation. For example, the coefficient of variation is 2.9\% (2.4/81.9) for the overall score, compared with 4.3\% (0.36/8.40) for the correctness score. The accounting papers (Panel B) exhibit similar patterns. And the authors are strong, as expected at these meetings: 56\% of the finance papers have an author at a top-twenty school, 82\% have an author with a prior top-three article, and 28\% have an author who has served on the editorial board of a top-three journal.

\section{Results}
\label{sec:results}

We proceed in seven steps. We first show that the model's verdict predicts publication and locate the information in its judgements of correctness and exposition (Sections~\ref{sec:results_main} and \ref{sec:results_dimensions}). We then show that a referee guide does not help (\autoref{sec:results_guidance}), that the pattern holds for accounting papers (\autoref{sec:results_acct}), and that a stronger model reproduces it (\autoref{sec:results_models}). The last two steps ask where the information comes from: not from the authors' standing or editorial connections (\autoref{sec:results_authors}), and not from memory of the outcomes (\autoref{sec:results_memory}).

\subsection{The LLM's Verdict Predicts Publication}
\label{sec:results_main}

Column~(1) of \autoref{tab:logit_main} relates the overall score to publication in a top-three finance journal. The coefficient is 0.495 ($z = 2.39$). Papers scored one standard deviation higher are 44\% more likely to be published in the \acs{jf}, \acs{jfe}, or \acs{rfs} within two years: 2.9 percentage points on a base rate of 6.5\% in the estimation sample.\footnote{The 2.9 percentage points are the average marginal effect of the standardized overall score in column~(1): for each of the 276 papers we compute the change in the predicted probability of publication per unit of the standardized score, with the controls at their observed values, and average over papers; because the score is standardized, a unit is one standard deviation. The delta-method standard error of the average marginal effect is 1.3 percentage points ($z = 2.16$). The base rate is the share of the 276 papers in a top-three journal, 18 papers or 6.5\% (\autoref{tab:outcomes}). The 44\% is the ratio of the two, 2.90 over 6.52.} The estimate is net of author reputation. Neither the authors' school, nor their prior top-three record, nor their editorial experience predicts publication in this sample. These papers have already passed a program committee's selection, which correlates with reputation. What remains to be explained is the editorial decision among papers of similar standing, and the model's reading of the paper carries information about that decision that the authors' names do not. Among the controls, only the number of authors predicts publication (Internet Appendix~\ref{sec:appendix_additional}).

The result is not specific to the top-three journals. Adding the \emph{Journal of Financial and Quantitative Analysis} and the \emph{Review of Finance} gives a coefficient of 0.451 on the overall score ($z = 2.39$), and adding the second tier of finance journals gives 0.561 ($z = 2.92$) (\autoref{tab:logit_fin_broad}).

\subsection{Right and Clear, Not Important}
\label{sec:results_dimensions}

Which judgement carries the information? Columns~(2) to (5) of \autoref{tab:logit_main} enter the four dimension scores one at a time. Correctness (0.691, $z = 2.00$) and exposition (0.667, $z = 2.68$) predict publication. Originality ($-0.069$, $z = -0.33$) and significance ($-0.141$, $z = -0.56$) carry no information: their coefficients are small, wrongly signed, and far from significant. The split is the same for the broader journal sets, where exposition again predicts on its own, correctness does so at the 10\% level, and originality and significance again carry no information (\autoref{tab:logit_fin_broad}).

The judgements of correctness and of contribution are separate in the model's scores. Across the finance papers, the correctness score is almost uncorrelated with the originality score (0.05) and the significance score ($-0.08$), and the same holds for the accounting papers ($-0.13$ and 0.01; \autoref{tab:dim_corr} in the Internet Appendix). A paper that the model rates as more correct is not rated as more original or more significant: the model makes two separate judgements, one of whether a paper is right and one of whether it matters. Its overall score is correlated with all four dimension scores (between 0.48 and 0.62 for the finance papers) and so combines the two.

For top-three publication, the model's overall score therefore dilutes its best signals. Correctness alone (pseudo-$R^2$ of 0.092) and exposition alone (0.096) explain more of the variation in publication than the overall score (0.072), which mixes these judgements with judgements of originality and significance that carry none.

The split runs against how the profession describes its own decisions. \citet{berk2017} advise referees that most submitted papers are correct, so that the decision often turns on the breadth of the question and the size of the advance over the literature, and that assessing importance is the hardest part of the job. The model's information lies in the dimensions that, by this account, most submissions already clear, and it has none in the dimensions on which the decision often turns. The pattern is consistent with what each judgement requires. Whether a design identifies what it claims, whether the evidence supports the interpretation, and whether the argument is clearly made can be checked against the paper itself. Whether a question matters, and whether the answer is new, call for a judgement about where the paper stands in a literature that keeps moving and about what the field values, and the model does not yet make this judgement as editors do. In the terms of \citet{ellison2002qr}, who separates the importance of a paper's main ideas from its other qualities such as rigour and polish, the model judges the second and not the first.

\subsection{A Referee Guide Does Not Help}
\label{sec:results_guidance}

A referee who uses a model can give it the advice that human referees follow, and a referee who wants the model to judge contribution might expect this to help. We test this on the same papers, keeping the model, the four draws, and the regression sample the same. The prompt places the full text of \citet{berk2017}, the advice summarized in \autoref{sec:results_dimensions}, before the naive prompt.

The guide does not help (\autoref{tab:logit_guide}). Compared with the naive prompt, it gives neither the originality nor the significance score any predictive power for top-three publication, and the information again lies in correctness (0.758, $z = 2.16$). Its overall score no longer predicts publication (0.203, $z = 0.80$), nor does its exposition score (0.007, $z = 0.03$).

The reading we favour is that the guide tells the model what to look for without giving it what it lacks: knowing that importance is what matters does not supply the judgement needed to assess importance.

\subsection{Accounting Papers}
\label{sec:results_acct}

Is the pattern specific to finance? Accounting provides a useful comparison because the two disciplines share confidential peer-review processes, selective publication standards, and a strong emphasis on research contributions. However, accounting research often requires specialized knowledge of reporting standards, auditing practices, and institutional settings. These differences may affect how \acsp{llm} assess research quality and the extent to which their evaluations align with editorial decisions. In addition, leading accounting journals have adopted particularly restrictive policies on referees' use of \acsp{llm} (\autoref{tab:llm_policies}), potentially limiting their adoption in the peer-review process. These considerations make accounting an interesting discipline for assessing the generalizability of our findings. We repeat the test of \autoref{tab:logit_main} on the 41 accounting papers described in \autoref{sec:data}, five of which are in a top-three accounting journal within two years. The model does not know these outcomes, so no accounting paper is dropped (columns~(3)--(4) of \autoref{tab:probes_desc}).

\autoref{tab:logit_acct} reports the estimates. The overall score predicts publication (1.558, $z = 2.23$), and the information again lies in correctness (1.506, $z = 2.09$) and exposition (1.271, $z = 1.94$) rather than in originality (0.904, $z = 1.62$) or significance ($-0.237$, $z = -0.37$). The sample is small and the coefficients are correspondingly large and imprecise, so we read the table for its pattern rather than its magnitudes. In accounting as in finance, the model's verdict carries information about the editorial decision.

\subsection{A Stronger Model}
\label{sec:results_models}

GPT-4o was released in 2024, and a natural question is whether a newer model closes the gap on originality and significance: a model that has read more of the literature might judge novelty and importance where its predecessor cannot. We repeat the design with GPT-5 (\texttt{gpt-5-2025-08-07}), which replaced GPT-4o as the default model of ChatGPT in August 2025, giving it the naive prompt. Its knowledge cutoff, 30 September 2024, falls after some of the publications we study, so we repeat with GPT-5 the look-ahead probe, which asks the model for the publication status of each paper (\autoref{sec:results_memory}). Given the title and authors, it names the correct journal of four finance papers. Two of them are known to have been submitted or online before its cutoff, and we drop these two from the sample of \autoref{tab:logit_main} and of this section.

GPT-5 grades more strictly than GPT-4o: it gives every paper a lower correctness score, by 1.74 points on average for the finance papers, and on average lower originality and significance scores, while its overall score for the finance papers is unchanged (\autoref{tab:gpt5_change} in the Internet Appendix). Because the scores are standardized, a stricter scale does not change the estimates.

GPT-5 reproduces the pattern (\autoref{tab:logit_gpt5}). Its information again lies in correctness: the correctness score predicts top-three publication (0.677, $z = 2.31$), with a pseudo-$R^2$ of 0.086 against 0.080 for the overall score. Its overall score predicts publication at the 10\% level (0.806, $z = 1.88$), where GPT-4o's does at the 5\% level on the same papers (0.495 in \autoref{tab:logit_main}). The exposition result is also robust to GPT-5 (0.490, $z = 1.71$). Neither originality ($-0.148$, $z = -0.66$) nor significance (0.389, $z = 1.33$) predicts publication.

The stronger model therefore does not change the answer. It is about as informative as its predecessor about whether a paper is right, and no more informative about whether it matters, although it has read more of the literature. Knowing more does not by itself supply the judgement that the decision on contribution requires.

\subsection{Author and Editor Information}
\label{sec:results_authors}

The conference PDF shows who wrote the paper, and the model reads the names, affiliations, and acknowledgements. Author standing predicts publication in the universe of finance working papers \citep{dai2023}, and a connection to an editor raises the chance of publication at the editor's journal \citep{brogaard2014}. If the model's score reflected these facts rather than the content of the paper, our conclusion would change: the model would be a reader of title pages, not of papers.

One result already speaks against this. The authors' school, prior top-three record, and editorial experience are controlled for in every specification, and none of them predicts publication in the conference sample (\autoref{sec:results_main}), so there is little for the score to proxy for. For comparison, human referees respond strongly to standing: in a field experiment with a finance manuscript, the share of reviewers recommending rejection fell from 65\% when only an unknown author was shown to 23\% when only a Nobel laureate was shown \citep{huber2022}. Interacting the score with indicators for an author with a prior top-three article, an editor among the authors, or an all-early-stage team, following Table~11 of \citet{dai2023}, gives no interaction that is significant at the 10\% level (\autoref{tab:logit_reputation} in the Internet Appendix).

Editorial connections are the sharper test, because an editor's name on a title page is a fact the model can read and a signal that an editor's decision might follow \citep{laband1994, brogaard2014}. ${\it Editorial}_i$ (\autoref{sec:data}) equals one for 28\% of the finance papers and 24\% of the accounting papers (\autoref{tab:sum_scores}), and it is insignificant in every column of Tables~\ref{tab:logit_main} to \ref{tab:logit_gpt5}, including those for the referee guide, the accounting papers, and GPT-5. On the evidence of the author characteristics we measure, the score's information is not explained by who wrote the paper.

\subsection{Not Memory}
\label{sec:results_memory}

Another alternative is that the model remembers the outcomes (look-ahead bias). Every publication in the sample appeared after the model's cutoff; the earliest went online on 28 October 2023. A signal could still have leaked through an author's CV listing a revise-and-resubmit. We therefore ask the model, in a fresh session, for the publication status of every paper. Given the title, it answers that it does not know for 96\% of the papers; given the title and authors, for 90\% (\autoref{tab:probes_desc}). It never reports a revise-and-resubmit, and its occasional claims that a paper is published are guesses, made more often for papers that remain unpublished than for those that are published. Adding an indicator for a claimed publication and its interaction with the score \citep{gao2025} leaves the coefficient on the score at 0.491 and 0.466 ($z = 2.29$ and 2.18) for the two information sets (Internet Appendix~\ref{sec:appendix_lookahead}). For the accounting papers the model answers that it does not know for every paper given the title and authors, and for all but one given the title alone (columns~(3) and (4) of \autoref{tab:probes_desc}).

\section{What the Results Imply for the Rules on AI in Peer Review}
\label{sec:implications}

Journals and conferences are writing rules for referees' use of \acsp{llm} without evidence on what the models can judge. \autoref{tab:llm_policies} collects the rules in force on 30 September 2026 at the leading computer science conferences, at the top finance and accounting journals and their publishers, and at two leading economics journals. A survey of 802 journals finds that most high-impact journals now have some rule of this kind \citep{wang2026policies}. Two patterns stand out in the table.

First, the fields disagree about \emph{whether}. In computer science, CVPR 2025, NeurIPS 2025, and the ACL Rolling Review ban or restrict reviewers' use of the models, ICML 2026 lets authors choose the policy for their paper, ICLR 2026 permits use with disclosure, and AAAI-26 and NeurIPS 2026 supply \acs{llm} output to their own review process. In finance, accounting, and economics, the rules run from permission under conditions of accountability, confidentiality, or disclosure at the \emph{Journal of Finance}, the \emph{Review of Financial Studies}, the \emph{Review of Finance}, the \emph{American Economic Review}, and \emph{Econometrica}, through bans on uploading a manuscript to an AI tool at the \emph{Journal of Financial Economics}, the \emph{Journal of Accounting Research}, and \emph{The Accounting Review}, or to a public tool at the \emph{Review of Accounting Studies} and \emph{Management Science}, to an outright ban at \emph{Contemporary Accounting Research}.

Second, where the rules permit, the permitted use is clerical. A referee may draft, edit, organize, and look up references; the assessment stays with the referee. The \emph{Review of Finance} says the models must not substitute for the referee's own evaluation, and \emph{Econometrica} that the referee must form an independent evaluation. \emph{Management Science} says that all conclusions must be the reviewer's own, and \emph{The Accounting Review} states that the critical assessment required for peer review is outside the scope of generative AI. The premise behind the permissive and the restrictive rules is the same: the model's verdict on a paper is not to be trusted. The rules differ only in whether the drafting help is worth the risk to confidentiality.

Our evidence refines that premise, and the refinement matters for how the rules should be written. The critical assessment of a paper has two parts. For the first part, whether the analysis is sound, the model's judgements of whether the analysis is sound and whether the paper is clearly written carry information about the editorial decision for GPT-4o, and the results are robust to the accounting papers and to GPT-5. For the second part, whether the question matters and the answer is new, the model's judgement carries no information, and asking it to weigh that part, as the referee guide does, does not help. A rule that treats the model's verdict as one thing therefore makes one of two mistakes. A ban discards a check that is informative on the margin of correctness. A deployment that passes on the model's overall score, whether to triage submissions or to seed a review, admits a judgement that mixes an informative part with an uninformative one, and for top-three publication in finance the mix predicts worse than the correctness score alone (\autoref{sec:results_dimensions}). Our findings suggest an important distinction between the tasks that \acsp{llm} can usefully support, such as assessing technical correctness and clarity, and those that require greater human judgement, such as evaluating a paper's intellectual contribution.

Some venues already separate the two tasks, which is itself evidence that the division is workable. AAAI-26's \acs{llm} reviews carried no scores or recommendations, and the authors and committee members who read them rated them above human reviews at identifying technical errors, the margin on which our evidence says the model is informative \citep{biswas2026aaai}. \emph{Econometrica} runs its conditionally accepted papers through an AI tool whose stated purpose is to identify technical and expository issues, not to assess the contribution of a paper, while requiring referees to form their own evaluation.\footnote{See \url{https://www.econometricsociety.org/publications/econometrica/editorial-procedures-and-policies} and \url{https://www.econometricsociety.org/publications/econometrica/guidelines-referees}.} \citet{bhargava2026fire} propose that journals place the models in a structured, contestable front-line review under human oversight rather than leave referees to use them privately. \citet{gans2026} shows that how an editor shares an AI assessment with referees determines whether it adds information or duplicates the model's blind spots. Our results say which assessment is worth sharing.

Two consequences follow for the venues in \autoref{tab:llm_policies}. For a referee working under a rule like that of the \emph{Journal of Finance}, the useful request is a technical check of the design, the evidence, and the writing, and the guide experiment shows that asking the model to apply the field's standard of contribution does not help. And the upload bans bind: the check requires the model to read the full paper, which a referee at the \emph{Journal of Financial Economics}, the \emph{Journal of Accounting Research}, \emph{The Accounting Review}, or \emph{Management Science} may not provide to a public tool. Under these rules the check can be run by the journal itself, under its own confidentiality terms, as the computer science conferences do, or by a referee through a tool that guarantees confidentiality, as the \emph{Review of Financial Studies} rule allows. It can also be run by the authors before submission, where no confidentiality question arises. Authors are the one party that no rule restricts, and the technical check is the one they can use. Bans on referees are in any case hard to enforce: in a randomized comparison at ICML 2026, a ban and a limited-use policy had near-zero effects on review scores and decisions, and 22.5\% of the reviewers under the ban reported using a model anyway \citep{kim2026icml}.

Two cautions apply. The first is that the anticipated use of \acs{llm} evaluations may induce authors to optimize their manuscripts for model-generated scores rather than for research quality. Firms revise their disclosures once they know that machines read them \citep{cao2023machine}, and authors have already hidden instructions to AI reviewers in the text of preprints \citep{gibney2025hidden}. A journal that runs the check must therefore treat its output as a reading to be verified rather than as a verdict, and keep the prompt and the model under its own control. The second is that our evidence has limits that the rules should respect. It comes from a naive prompt and two models, it compares the model's verdict with the decision rather than with the reports that editors read, and it concerns papers that had already passed a conference's selection. It does not show that a model can replace the technical part of a referee report, only that its judgement on that part is informative. A journal that runs the check will want to measure, on its own submissions and reports, how much of the information survives. Only the journal can do this, because it alone holds the reports.

\section{Conclusion}
\label{sec:conclusion}

Which part of a referee's job can a large language model do? We answer with the one public record of refereeing in finance: the publication decision. Given three sentences of instruction, GPT-4o's verdict on the papers of the 2024 \acs{afa} and \acs{nber} meetings predicts which of them the top-three finance journals published within two years, net of the authors' reputation and with no memory of the outcomes. A one-standard-deviation rise in its score raises the chance of publication by 44\%. The prediction rests on two of the four judgements we ask for, correctness and exposition. The other two, originality and significance, add nothing, and a referee guide that tells the model to weigh importance does not help. The accounting papers and GPT-5 give the same split between correctness and contribution.

The results answer three questions that finance and accounting researchers are asking. Can a model replace a referee? No: under the naive prompt its judgements of originality and significance carry no information on the margin of contribution, which is the margin on which editors say their decisions turn. Can it assist human reviewers? Yes: its judgements of whether a paper is right and whether it is clearly written carry information for GPT-4o, and the results are robust to the accounting papers and to GPT-5. This is useful to referees drafting a report, to editors facing a growing volume of submissions, and to authors before they submit. What should the rules say? The technical check can go to the model, and the judgement of contribution should stay with people. Judgements of contribution remain inherently contextual and are not readily captured by model-generated scores, so at this stage our evidence suggests using \acsp{llm} as complementary tools in peer review while preserving human judgement as central to evaluating contributions.

Unlike some leading computer science conferences, finance and accounting journals keep their referee reports confidential. An \acs{llm} referee for our fields can therefore be neither calibrated on the judgements of our referees nor evaluated against them, and we can compare its verdicts only with publication decisions, a coarser signal that arrives years later. A release of anonymized referee reports and decisions by finance and accounting journals would allow researchers to compare a model's reports with those that editors read, to calibrate referees for our fields, and to measure which parts of human judgement the models can and cannot yet reproduce. A future step is to move from the naive prompt studied here to the rubric-guided design of \citet{newton2026}, and to measure whether the input of human judgement makes \acs{llm} scores of originality and significance predict better.

\newpage
\clearpage 
\bibliographystyle{chicago}
\bibliography{references}


\clearpage
{\footnotesize\renewcommand{\baselinestretch}{1}\selectfont
\begin{xltabular}{\linewidth}{@{}>{\raggedright\arraybackslash}p{0.2\linewidth} c X l@{}}
    \caption{\textbf{Rules for Large Language Models in Peer Review}\\This table summarizes the rules for reviewers' use of large language models at leading computer science conferences (Panel A), at finance and accounting journals and their publishers (Panel B), and at economics and management journals (Panel C), from the venues' own policy pages accessed 30 September 2026. Year is the conference year or, for journals, the year of the policy page. A publisher named in parentheses means that the journal states no rule of its own and the publisher's rule applies. ``Bans'' and ``Restricts'' mark rules that forbid or limit the use of LLMs by reviewers; ``Permits'' marks rules that allow it under conditions; ``Choice'' marks a rule under which authors choose the policy for their paper and reviewers are assigned one; ``Deploys'' marks venues that themselves supply LLM output to their review process.}
    \label{tab:llm_policies} \\
    \toprule
    Venue & Year & Rule for reviewers & Stance \\
    \midrule
    \endfirsthead
    \multicolumn{4}{@{}l}{\textit{Table \thetable\ continued}} \\
    \toprule
    Venue & Year & Rule for reviewers & Stance \\
    \midrule
    \endhead
    \midrule
    \multicolumn{4}{@{}r@{}}{\textit{Continued on next page}} \\
    \endfoot
    \midrule
    \multicolumn{4}{@{}l@{}}{\textit{Sources, in table order (all pages accessed 30 September 2026):}} \\
    \multicolumn{4}{@{}p{\linewidth}@{}}{CVPR, 2025 Reviewer Guidelines, \url{https://cvpr.thecvf.com/Conferences/2025/ReviewerGuidelines}} \\
    \multicolumn{4}{@{}p{\linewidth}@{}}{NeurIPS, 2025 Policy on the Use of Large Language Models, \url{https://neurips.cc/Conferences/2025/LLM}} \\
    \multicolumn{4}{@{}p{\linewidth}@{}}{ACL Rolling Review, Reviewer Guidelines, \url{https://aclrollingreview.org/reviewerguidelines}} \\
    \multicolumn{4}{@{}p{\linewidth}@{}}{ICML, 2026 Policy for LLM Use in Reviewing, \url{https://icml.cc/Conferences/2026/LLM-Policy}} \\
    \multicolumn{4}{@{}p{\linewidth}@{}}{ICLR, ICLR 2026 Response to LLM-Generated Papers and Reviews, blog post of 19 November 2025, \url{https://blog.iclr.cc/2025/11/19/iclr-2026-response-to-llm-generated-papers-and-reviews/}} \\
    \multicolumn{4}{@{}p{\linewidth}@{}}{ICLR, A Retrospective on the ICLR 2026 Review Process, blog post of 31 March 2026, \url{https://blog.iclr.cc/2026/03/31/a-retrospective-on-the-iclr-2026-review-process/}} \\
    \multicolumn{4}{@{}p{\linewidth}@{}}{AAAI, AAAI Launches AI-Powered Peer Review Assessment System, 16 May 2025, \url{https://aaai.org/aaai-launches-ai-powered-peer-review-assessment-system/}} \\
    \multicolumn{4}{@{}p{\linewidth}@{}}{NeurIPS, 2026 AI-Assisted Reviewing Experiment, \url{https://neurips.cc/Conferences/2026/ai-reviewing-experiment}} \\
    \multicolumn{4}{@{}p{\linewidth}@{}}{American Finance Association, Journal of Finance submissions page, \url{https://afajof.org/submissions/}} \\
    \multicolumn{4}{@{}p{\linewidth}@{}}{Society for Financial Studies, Review of Financial Studies: Submit a Report, \url{https://sfs.org/review-of-financial-studies/submit-a-report/}} \\
    \multicolumn{4}{@{}p{\linewidth}@{}}{Review of Finance, Policy Update Regarding the Use of Large Language Models, 23 March 2026, \url{https://revfin.org/policy-update-regarding-the-use-of-large-language-models-llms/}} \\
    \multicolumn{4}{@{}p{\linewidth}@{}}{Elsevier, Generative AI Policies for Journals, updated June 2026, \url{https://www.elsevier.com/about/policies-and-standards/generative-ai-policies-for-journals}} \\
    \multicolumn{4}{@{}p{\linewidth}@{}}{American Accounting Association, The Accounting Review: Guide for Authors, \url{https://aaahq.org/Research/Journals/The-Accounting-Review/Guide-for-Authors}} \\
    \multicolumn{4}{@{}p{\linewidth}@{}}{Wiley, Using AI Tools in Your Research, \url{https://www.wiley.com/en-us/publish/article/ai-guidelines/}} \\
    \multicolumn{4}{@{}p{\linewidth}@{}}{Springer Nature, Editorial Policies: Using AI in Peer Review, \url{https://www.springernature.com/gp/policies/editorial-policies/using-ai-in-peer-review}} \\
    \multicolumn{4}{@{}p{\linewidth}@{}}{Canadian Academic Accounting Association, Contemporary Accounting Research Editorial Policies, \url{https://www.caaa.ca/car-editorial-policies}} \\
    \multicolumn{4}{@{}p{\linewidth}@{}}{American Economic Association, American Economic Review Editorial Policy, \url{https://www.aeaweb.org/journals/aer/editorial-policy}} \\
    \multicolumn{4}{@{}p{\linewidth}@{}}{Econometric Society, Econometrica Reviewer Guidelines, \url{https://www.econometricsociety.org/publications/econometrica/guidelines-referees}} \\
    \multicolumn{4}{@{}p{\linewidth}@{}}{Econometric Society, Econometrica Editorial Procedures and Policies, \url{https://www.econometricsociety.org/publications/econometrica/editorial-procedures-and-policies}} \\
    \multicolumn{4}{@{}p{\linewidth}@{}}{INFORMS, Management Science Submission Guidelines, \url{https://pubsonline.informs.org/page/mnsc/submission-guidelines}} \\
    \bottomrule
    \endlastfoot
\multicolumn{4}{@{}l}{\textit{Panel A: Computer science}} \\*[2pt]
CVPR & 2025 & LLMs may not be used to write reviews or meta-reviews at any step; a violating reviewer's own papers may be desk rejected & Bans \\
NeurIPS & 2025 & Submissions may not be shared with any LLM; LLMs may help a reviewer understand concepts and improve the writing & Restricts \\
ACL Rolling Review & 2026 & Generative assistance may not produce the first draft of a review; paraphrasing is allowed, uploads to tools that do not preserve privacy are not & Restricts \\
ICML & 2026 & Authors state whether their paper requires a no-LLM policy or allows a permissive one, and each reviewer is assigned one policy; the permissive policy lets an LLM explain the paper and polish the review but not judge it & Choice \\
ICLR & 2026 & Reviewers may use LLMs with disclosure and are responsible for the output; reviews with false claims or hallucinated references breach the code of ethics, and the reviewer's own papers may be desk rejected; all reviews were screened with two LLM detectors and the flags sent to area chairs & Permits \\
AAAI & 2026 & The conference adds LLM-generated reviews to the initial review stage alongside human reviews, without numerical scores or ratings and not for automated accept or reject decisions; no human reviewer is replaced & Deploys \\
NeurIPS & 2026 & A voluntary randomized experiment gives reviewers an LLM assistant in OpenReview that helps them analyse a submission but does not write the review; outside the experiment no LLM use is sanctioned & Deploys \\
\midrule
\multicolumn{4}{@{}l}{\textit{Panel B: Finance and accounting}} \\*[2pt]
\emph{Journal of Finance} & 2026 & AI may help draft or edit referee reports and look up prior references; referees must fact-check its output & Permits \\
\emph{Rev.\ Fin.\ Studies} & 2026 & Manuscripts may not be fed into any tool that does not guarantee strict confidentiality, including an opt-out from model training; the referee remains responsible for the report & Permits \\
\emph{Review of Finance} & 2026 & LLMs may be used in a supportive way but not as a substitute for the referee's own evaluation, interpretation, or recommendations & Permits \\
\emph{J.\ Financial Economics} (Elsevier) & 2026 & Reviewers should not upload a manuscript or any part of it into an AI tool; private tools may support language and literature search, with disclosure & Restricts \\
\emph{Accounting Review} & 2026 & Reviewers should not upload any part of a manuscript into an AI tool; AI may improve the language of the report, with disclosure; the critical assessment required for peer review is outside the scope of generative AI & Restricts \\
\emph{J.\ Accounting Research} (Wiley) & 2026 & Manuscripts may not be uploaded to AI tools, in full or in part; AI drafting assistance must be disclosed & Restricts \\
\emph{Rev.\ Acc.\ Studies} (Springer Nature) & 2026 & Manuscript content may not be uploaded to public AI tools; generating a review, producing a recommendation, or delegating critique to AI is not permitted & Restricts \\
\emph{Contemp.\ Acc.\ Research} & 2026 & Reviewers are prohibited from using generative AI tools in the review process, including to improve the language of a completed report & Bans \\
\midrule
\multicolumn{4}{@{}l}{\textit{Panel C: Economics and management}} \\*[2pt]
\emph{Amer.\ Economic Review} & 2026 & Any AI use must be disclosed in the reviewer's cover letter and limited to tools that guarantee the manuscript will not be used for training & Permits \\
\emph{Econometrica} & 2026 & Referees are expected to read the submission themselves and form an independent evaluation; AI may improve the clarity, organization, and language of the report; conditionally accepted papers are checked with an AI tool for technical and expository issues, not for contribution & Permits \\
\emph{Management Science} & 2026 & No part of a manuscript may be uploaded to a generative AI tool that could compromise confidentiality; all conclusions must be the reviewer's own & Restricts \\
\end{xltabular}}

\clearpage
\begin{table}[H]
    \caption{\textbf{Summary Statistics}\\This table presents summary statistics for the 276 papers presented at the 2024 AFA Annual Meeting and the six 2024 NBER finance meetings outside the AFA PhD student poster session that enter the regressions (Panel A; \autoref{sec:spec}) and for the 41 accounting papers presented at the 2023 and 2024 Conferences on Financial Economics and Accounting (Panel B; Internet Appendix~\ref{sec:appendix_acct}). Scores are the means over four independent draws of GPT-4o (\texttt{gpt-4o-2024-11-20}) under the naive prompt reproduced in Internet Appendix~\ref{sec:prompt}, on a 0--100 scale for the overall score and a 0--10 scale for the four dimensions. Publication outcomes are measured within two years of each paper's meeting. In Panel B, ${\it Log\ Prior\ Top}_i$ counts the authors' articles in the three accounting journals of ${\it Top3Acc}_i$ before 2024, ${\it Editorial}_i$ refers to the same journals, and ${\it Top20School}_i$ uses the same list of twenty finance schools as Panel A. All variables are defined in \autoref{tab:var_des}.}
    \small
    \centering
    \begingroup\renewcommand{\baselinestretch}{1}\selectfont
    \textit{Panel A: Finance}\\[2pt]
    \begin{tabular}{l*{6}{c}}
\toprule
 & N & Mean & SD & P10 & P50 & P90 \\
\midrule
${\it Overall}_i$ & 276  &  81.924  &  2.402  &  78.75  &  82.25  &  85.00 \\
${\it Originality}_i$ & 276  &  8.308  &  0.310  &  8.00  &  8.25  &  8.75 \\
${\it Significance}_i$ & 276  &  8.118  &  0.370  &  7.50  &  8.00  &  8.50 \\
${\it Correctness}_i$ & 276  &  8.397  &  0.363  &  8.00  &  8.50  &  8.75 \\
${\it Exposition}_i$ & 276  &  7.207  &  0.410  &  6.75  &  7.25  &  7.75 \\
${\it Top3Fin}_i$ & 276  &  0.065  &  0.247  &  0.00  &  0.00  &  0.00 \\
${\it Top5Fin}_i$ & 276  &  0.072  &  0.260  &  0.00  &  0.00  &  0.00 \\
${\it Top10Fin}_i$ & 276  &  0.080  &  0.271  &  0.00  &  0.00  &  0.00 \\
${\it Log\ Length}_i$ & 276  &  10.396  &  0.351  &  9.96  &  10.42  &  10.82 \\
${\it N\ Authors}_i$ & 276  &  2.931  &  0.964  &  2.00  &  3.00  &  4.00 \\
${\it NBER}_i$ & 276  &  0.167  &  0.373  &  0.00  &  0.00  &  1.00 \\
${\it Top20School}_i$ & 276  &  0.558  &  0.498  &  0.00  &  1.00  &  1.00 \\
${\it Established}_i$ & 276  &  0.822  &  0.383  &  0.00  &  1.00  &  1.00 \\
${\it Log\ Prior\ Top}_i$ & 276  &  1.730  &  1.102  &  0.00  &  1.95  &  3.04 \\
${\it Early\ Stage}_i$ & 276  &  0.109  &  0.312  &  0.00  &  0.00  &  1.00 \\
${\it Editorial}_i$ & 276  &  0.275  &  0.448  &  0.00  &  0.00  &  1.00 \\
\bottomrule
\end{tabular}

    \vspace{8pt}
    \textit{Panel B: Accounting}\\[2pt]
    \begin{tabular}{l*{6}{c}}
\toprule
 & N & Mean & SD & P10 & P50 & P90 \\
\midrule
${\it Overall}_i$ & 41  &  82.537  &  2.252  &  79.50  &  82.75  &  85.00 \\
${\it Originality}_i$ & 41  &  8.183  &  0.262  &  8.00  &  8.25  &  8.50 \\
${\it Significance}_i$ & 41  &  8.061  &  0.402  &  7.50  &  8.00  &  8.50 \\
${\it Correctness}_i$ & 41  &  8.506  &  0.401  &  8.00  &  8.50  &  9.00 \\
${\it Exposition}_i$ & 41  &  7.232  &  0.360  &  6.75  &  7.25  &  7.75 \\
${\it Top3Acc}_i$ & 41  &  0.122  &  0.331  &  0.00  &  0.00  &  1.00 \\
${\it Log\ Length}_i$ & 41  &  10.334  &  0.212  &  10.10  &  10.31  &  10.69 \\
${\it N\ Authors}_i$ & 41  &  2.732  &  1.096  &  1.00  &  3.00  &  4.00 \\
${\it Top20School}_i$ & 41  &  0.341  &  0.480  &  0.00  &  0.00  &  1.00 \\
${\it Log\ Prior\ Top}_i$ & 41  &  1.631  &  0.972  &  0.00  &  1.79  &  2.64 \\
${\it Editorial}_i$ & 41  &  0.244  &  0.435  &  0.00  &  0.00  &  1.00 \\
\bottomrule
\end{tabular}

    \endgroup
    \label{tab:sum_scores}
\end{table}

\clearpage
\begin{table}[H]
    \centering
    \small
    \caption{\textbf{The LLM Referee's Scores and Publication in the Top Three Finance Journals}\\This table presents logit estimates of equation~\eqref{eq:logit}. The sample is the 276 papers presented at the 2024 AFA Annual Meeting and the six 2024 NBER finance meetings outside the AFA PhD student poster session that enter the regressions: of the 280 papers outside the poster session, we drop the paper none of whose authors could be matched to OpenAlex, one paper for which GPT-4o names the journal of an article submitted before its knowledge cutoff (Internet Appendix~\ref{sec:appendix_lookahead}), and two papers whose journal GPT-5 names and that were submitted or online before its cutoff (\autoref{sec:results_models}). The outcome is ${\it Top3Fin}_i$, one if the paper is online at the \emph{Journal of Finance}, the \emph{Journal of Financial Economics}, or the \emph{Review of Financial Studies} within two years of its meeting. Scores are the means over four independent draws of GPT-4o (\texttt{gpt-4o-2024-11-20}) under the naive prompt reproduced in Internet Appendix~\ref{sec:prompt}, standardized across papers; each column enters one score, and a coefficient is the change in log odds per standard deviation of the score. All specifications include the log length of the paper text, the number of authors, ${\it Top20School}_i$, ${\it Log\ Prior\ Top}_i$, ${\it Editorial}_i$, and a conference (AFA versus NBER) fixed effect. All variables are defined in \autoref{tab:var_des}. Robust $z$-statistics in parentheses. ***, **, and * denote significance at the 1\%, 5\%, and 10\% levels.}
    \begingroup\renewcommand{\baselinestretch}{1}\selectfont
    {
\def\sym#1{\ifmmode^{#1}\else\(^{#1}\)\fi}
\begin{tabular}{l*{5}{c}}
\toprule
                    &\multicolumn{5}{c}{${\it Top3Fin}_i$}                                                                        \\\cmidrule(lr){2-6}
                    &\multicolumn{1}{c}{(1)}         &\multicolumn{1}{c}{(2)}         &\multicolumn{1}{c}{(3)}         &\multicolumn{1}{c}{(4)}         &\multicolumn{1}{c}{(5)}         \\
\midrule
${\it Overall}_i$   &       0.495\sym{**} &                     &                     &                     &                     \\
                    &      (2.39)         &                     &                     &                     &                     \\
${\it Originality}_i$&                     &      -0.069         &                     &                     &                     \\
                    &                     &     (-0.33)         &                     &                     &                     \\
${\it Significance}_i$&                     &                     &      -0.141         &                     &                     \\
                    &                     &                     &     (-0.56)         &                     &                     \\
${\it Correctness}_i$&                     &                     &                     &       0.691\sym{**} &                     \\
                    &                     &                     &                     &      (2.00)         &                     \\
${\it Exposition}_i$&                     &                     &                     &                     &       0.667\sym{***}\\
                    &                     &                     &                     &                     &      (2.68)         \\
${\it Log\ Length}_i$&       0.017         &       0.061         &       0.116         &      -0.050         &      -0.328         \\
                    &      (0.03)         &      (0.11)         &      (0.20)         &     (-0.08)         &     (-0.58)         \\
${\it N\ Authors}_i$&       0.597\sym{**} &       0.567\sym{**} &       0.573\sym{**} &       0.622\sym{**} &       0.623\sym{**} \\
                    &      (2.45)         &      (2.31)         &      (2.32)         &      (2.48)         &      (2.53)         \\
${\it Top20School}_i$&       0.171         &       0.261         &       0.257         &       0.231         &       0.133         \\
                    &      (0.34)         &      (0.51)         &      (0.53)         &      (0.48)         &      (0.27)         \\
${\it Log\ Prior\ Top}_i$&       0.108         &       0.086         &       0.097         &       0.130         &       0.113         \\
                    &      (0.39)         &      (0.31)         &      (0.35)         &      (0.50)         &      (0.37)         \\
${\it Editorial}_i$ &      -0.063         &       0.034         &       0.039         &      -0.159         &      -0.026         \\
                    &     (-0.11)         &      (0.06)         &      (0.07)         &     (-0.28)         &     (-0.05)         \\
\midrule
Conference FE       &           Y         &           Y         &           Y         &           Y         &           Y         \\
N Obs               &         276         &         276         &         276         &         276         &         276         \\
Pseudo R$^2$        &       0.072         &       0.048         &       0.050         &       0.092         &       0.096         \\
\bottomrule
\end{tabular}
}

    \endgroup
    \label{tab:logit_main}
\end{table}

\clearpage
\begin{table}[H]
    \centering
    \small
    \caption{\textbf{The Referee Guide and Publication in the Top Three Finance Journals}\\This table presents logit estimates of equation~\eqref{eq:logit} with scores obtained under a prompt that places the full text of the refereeing advice of \citet{berk2017} before the naive prompt (\autoref{sec:results_guidance}; both prompts are reproduced in Internet Appendix~\ref{sec:prompt}). The sample is the 276 papers presented at the 2024 AFA Annual Meeting and the six 2024 NBER finance meetings outside the AFA PhD student poster session that enter the regressions (\autoref{sec:spec}). The outcome is ${\it Top3Fin}_i$, one if the paper is online at the \emph{Journal of Finance}, the \emph{Journal of Financial Economics}, or the \emph{Review of Financial Studies} within two years of its meeting. Scores are the means over four independent draws of GPT-4o (\texttt{gpt-4o-2024-11-20}) under this guided prompt, standardized across papers; each column enters one score, and a coefficient is the change in log odds per standard deviation of the score. All specifications include the log length of the paper text, the number of authors, ${\it Top20School}_i$, ${\it Log\ Prior\ Top}_i$, ${\it Editorial}_i$, and a conference (AFA versus NBER) fixed effect. All variables are defined in \autoref{tab:var_des}. Robust $z$-statistics in parentheses. ***, **, and * denote significance at the 1\%, 5\%, and 10\% levels.}
    \begingroup\renewcommand{\baselinestretch}{1}\selectfont
    {
\def\sym#1{\ifmmode^{#1}\else\(^{#1}\)\fi}
\begin{tabular}{l*{5}{c}}
\toprule
                    &\multicolumn{5}{c}{Referee guide attached}                                                                   \\\cmidrule(lr){2-6}
                    &\multicolumn{1}{c}{(1)}         &\multicolumn{1}{c}{(2)}         &\multicolumn{1}{c}{(3)}         &\multicolumn{1}{c}{(4)}         &\multicolumn{1}{c}{(5)}         \\
\midrule
${\it Overall}_i$   &       0.203         &                     &                     &                     &                     \\
                    &      (0.80)         &                     &                     &                     &                     \\
${\it Originality}_i$&                     &      -0.189         &                     &                     &                     \\
                    &                     &     (-0.76)         &                     &                     &                     \\
${\it Significance}_i$&                     &                     &      -0.228         &                     &                     \\
                    &                     &                     &     (-0.97)         &                     &                     \\
${\it Correctness}_i$&                     &                     &                     &       0.758\sym{**} &                     \\
                    &                     &                     &                     &      (2.16)         &                     \\
${\it Exposition}_i$&                     &                     &                     &                     &       0.007         \\
                    &                     &                     &                     &                     &      (0.03)         \\
${\it Log\ Length}_i$&       0.030         &       0.103         &       0.041         &      -0.007         &       0.058         \\
                    &      (0.05)         &      (0.18)         &      (0.08)         &     (-0.01)         &      (0.10)         \\
${\it N\ Authors}_i$&       0.562\sym{**} &       0.579\sym{**} &       0.562\sym{**} &       0.638\sym{***}&       0.566\sym{**} \\
                    &      (2.35)         &      (2.30)         &      (2.27)         &      (2.81)         &      (2.31)         \\
${\it Top20School}_i$&       0.221         &       0.260         &       0.239         &       0.189         &       0.241         \\
                    &      (0.45)         &      (0.52)         &      (0.49)         &      (0.39)         &      (0.49)         \\
${\it Log\ Prior\ Top}_i$&       0.083         &       0.096         &       0.115         &       0.068         &       0.091         \\
                    &      (0.29)         &      (0.35)         &      (0.42)         &      (0.23)         &      (0.32)         \\
${\it Editorial}_i$ &       0.056         &       0.013         &      -0.024         &      -0.014         &       0.035         \\
                    &      (0.10)         &      (0.02)         &     (-0.04)         &     (-0.02)         &      (0.06)         \\
\midrule
Conference FE       &           Y         &           Y         &           Y         &           Y         &           Y         \\
N Obs               &         276         &         276         &         276         &         276         &         276         \\
Pseudo R$^2$        &       0.051         &       0.052         &       0.053         &       0.105         &       0.048         \\
\bottomrule
\end{tabular}
}

    \endgroup
    \label{tab:logit_guide}
\end{table}

\clearpage
\begin{table}[H]
    \centering
    \small
    \caption{\textbf{Accounting Papers: The LLM Referee's Scores and Publication in the Top Three Accounting Journals}\\This table presents logit estimates of equation~\eqref{eq:logit} for the 41 accounting papers presented at the 2023 and 2024 Conferences on Financial Economics and Accounting (Internet Appendix~\ref{sec:appendix_acct}). The outcome is ${\it Top3Acc}_i$, one if the paper is online at \emph{The Accounting Review}, the \emph{Journal of Accounting Research}, or the \emph{Journal of Accounting and Economics} within two years of its meeting. Scores are the means over four independent draws of GPT-4o (\texttt{gpt-4o-2024-11-20}) under the naive prompt reproduced in Internet Appendix~\ref{sec:prompt}, addressed to an accounting journal, standardized across papers; each column enters one score, and a coefficient is the change in log odds per standard deviation of the score. All specifications include the log length of the paper text, the number of authors, ${\it Top20School}_i$ (the same list of twenty finance schools as for the finance papers), ${\it Log\ Prior\ Top}_i$, which counts the authors' articles published before 2024 in the three accounting journals, ${\it Editorial}_i$ for the same journals, and a fixed effect for the meeting (2023 versus 2024). All variables are defined in \autoref{tab:var_des}. Robust $z$-statistics in parentheses. ***, **, and * denote significance at the 1\%, 5\%, and 10\% levels.}
    \begingroup\renewcommand{\baselinestretch}{1}\selectfont
    {
\def\sym#1{\ifmmode^{#1}\else\(^{#1}\)\fi}
\begin{tabular}{l*{5}{c}}
\toprule
                    &\multicolumn{5}{c}{${\it Top3Acc}_i$}                                                                        \\\cmidrule(lr){2-6}
                    &\multicolumn{1}{c}{(1)}         &\multicolumn{1}{c}{(2)}         &\multicolumn{1}{c}{(3)}         &\multicolumn{1}{c}{(4)}         &\multicolumn{1}{c}{(5)}         \\
\midrule
${\it Overall}_i$   &       1.558\sym{**} &                     &                     &                     &                     \\
                    &      (2.23)         &                     &                     &                     &                     \\
${\it Originality}_i$&                     &       0.904         &                     &                     &                     \\
                    &                     &      (1.62)         &                     &                     &                     \\
${\it Significance}_i$&                     &                     &      -0.237         &                     &                     \\
                    &                     &                     &     (-0.37)         &                     &                     \\
${\it Correctness}_i$&                     &                     &                     &       1.506\sym{**} &                     \\
                    &                     &                     &                     &      (2.09)         &                     \\
${\it Exposition}_i$&                     &                     &                     &                     &       1.271\sym{*}  \\
                    &                     &                     &                     &                     &      (1.94)         \\
${\it Log\ Length}_i$&      -2.189         &       0.955         &       0.773         &      -2.256         &       1.035         \\
                    &     (-0.80)         &      (0.42)         &      (0.35)         &     (-0.91)         &      (0.36)         \\
${\it N\ Authors}_i$&      -0.864         &      -0.947         &      -0.915\sym{*}  &      -0.814         &      -1.051         \\
                    &     (-1.36)         &     (-1.18)         &     (-1.79)         &     (-1.38)         &     (-1.58)         \\
${\it Top20School}_i$&       1.637         &       1.505         &       1.674         &       1.070         &       1.693         \\
                    &      (1.17)         &      (1.26)         &      (1.34)         &      (0.80)         &      (1.34)         \\
${\it Log\ Prior\ Top}_i$&      -0.837         &      -0.143         &      -0.011         &      -0.447         &       0.015         \\
                    &     (-0.63)         &     (-0.15)         &     (-0.01)         &     (-0.47)         &      (0.01)         \\
${\it Editorial}_i$ &      -0.191         &      -0.667         &      -0.328         &      -0.356         &      -0.664         \\
                    &     (-0.07)         &     (-0.40)         &     (-0.18)         &     (-0.16)         &     (-0.41)         \\
\midrule
Conference FE       &           Y         &           Y         &           Y         &           Y         &           Y         \\
N Obs               &          41         &          41         &          41         &          41         &          41         \\
Pseudo R$^2$        &       0.280         &       0.207         &       0.163         &       0.248         &       0.301         \\
\bottomrule
\end{tabular}
}

    \endgroup
    \label{tab:logit_acct}
\end{table}

\clearpage
\begin{table}[H]
    \centering
    \small
    \caption{\textbf{GPT-5's Scores and Publication in the Top Three Finance Journals}\\This table presents logit estimates of equation~\eqref{eq:logit} with the scores of GPT-5 (\autoref{sec:results_models}). The sample is the 276 papers presented at the 2024 AFA Annual Meeting and the six 2024 NBER finance meetings outside the AFA PhD student poster session that enter the regressions (\autoref{sec:spec}). The outcome is ${\it Top3Fin}_i$, one if the paper is online at the \emph{Journal of Finance}, the \emph{Journal of Financial Economics}, or the \emph{Review of Financial Studies} within two years of its meeting. Scores are the means over four independent draws of GPT-5 (\texttt{gpt-5-2025-08-07}) under the naive prompt reproduced in Internet Appendix~\ref{sec:prompt}, standardized across papers; each column enters one score, and a coefficient is the change in log odds per standard deviation of the score. All specifications include the log length of the paper text, the number of authors, ${\it Top20School}_i$, ${\it Log\ Prior\ Top}_i$, ${\it Editorial}_i$, and a conference (AFA versus NBER) fixed effect. All variables are defined in \autoref{tab:var_des}. Robust $z$-statistics in parentheses. ***, **, and * denote significance at the 1\%, 5\%, and 10\% levels.}
    \begingroup\renewcommand{\baselinestretch}{1}\selectfont
    {
\def\sym#1{\ifmmode^{#1}\else\(^{#1}\)\fi}
\begin{tabular}{l*{5}{c}}
\toprule
                    &\multicolumn{5}{c}{${\it Top3Fin}_i$}                                                                        \\\cmidrule(lr){2-6}
                    &\multicolumn{1}{c}{(1)}         &\multicolumn{1}{c}{(2)}         &\multicolumn{1}{c}{(3)}         &\multicolumn{1}{c}{(4)}         &\multicolumn{1}{c}{(5)}         \\
\midrule
${\it Overall}_i$   &       0.806\sym{*}  &                     &                     &                     &                     \\
                    &      (1.88)         &                     &                     &                     &                     \\
${\it Originality}_i$&                     &      -0.148         &                     &                     &                     \\
                    &                     &     (-0.66)         &                     &                     &                     \\
${\it Significance}_i$&                     &                     &       0.389         &                     &                     \\
                    &                     &                     &      (1.33)         &                     &                     \\
${\it Correctness}_i$&                     &                     &                     &       0.677\sym{**} &                     \\
                    &                     &                     &                     &      (2.31)         &                     \\
${\it Exposition}_i$&                     &                     &                     &                     &       0.490\sym{*}  \\
                    &                     &                     &                     &                     &      (1.71)         \\
${\it Log\ Length}_i$&      -0.803         &       0.256         &      -0.296         &      -0.671         &      -0.263         \\
                    &     (-1.11)         &      (0.42)         &     (-0.46)         &     (-0.94)         &     (-0.44)         \\
${\it N\ Authors}_i$&       0.604\sym{**} &       0.556\sym{**} &       0.581\sym{**} &       0.611\sym{**} &       0.522\sym{**} \\
                    &      (2.39)         &      (2.31)         &      (2.36)         &      (2.46)         &      (2.12)         \\
${\it Top20School}_i$&      -0.019         &       0.270         &       0.111         &       0.023         &       0.103         \\
                    &     (-0.03)         &      (0.53)         &      (0.21)         &      (0.04)         &      (0.20)         \\
${\it Log\ Prior\ Top}_i$&       0.136         &       0.066         &       0.115         &       0.139         &       0.096         \\
                    &      (0.46)         &      (0.23)         &      (0.41)         &      (0.47)         &      (0.33)         \\
${\it Editorial}_i$ &      -0.125         &       0.093         &      -0.075         &      -0.171         &      -0.008         \\
                    &     (-0.21)         &      (0.16)         &     (-0.12)         &     (-0.28)         &     (-0.01)         \\
\midrule
Conference FE       &           Y         &           Y         &           Y         &           Y         &           Y         \\
N Obs               &         276         &         276         &         276         &         276         &         276         \\
Pseudo R$^2$        &       0.080         &       0.050         &       0.060         &       0.086         &       0.067         \\
\bottomrule
\end{tabular}
}

    \endgroup
    \label{tab:logit_gpt5}
\end{table}

\clearpage

\newpage
\newpage
\clearpage
\setcounter{section}{0}
\setcounter{table}{0}
\renewcommand{\thesection}{IA.\arabic{section}}
\renewcommand{\theHsection}{IA.\arabic{section}}
\renewcommand{\thetable}{IA.\arabic{table}}
\renewcommand{\theHtable}{IA.\arabic{table}}

\begin{center}
{\LARGE Internet Appendix}\\[8pt]
{\large for ``How Well Do LLMs Perform in Peer Review? Evidence from Finance and Accounting''}
\end{center}
\bigskip

\section{Sample and Text}
\label{sec:appendix_sample}

The finance sample consists of the papers presented at the 2024 \acs{afa} Annual Meeting (San Antonio, 5--7 January 2024) and at the six 2024 \acs{nber} finance meetings: the Asset Pricing and Corporate Finance program meetings in spring and fall, and the two Summer Institute meetings. We do not substitute repository versions for the conferences whose programs do not host the full text (\autoref{sec:data}).

We crawl the two programs and download every PDF the conference hosts. The \acs{afa} program lists 334 papers (243 in regular sessions and 91 in the PhD poster session) and hosts a PDF for 333 of them; the \acs{nber} programs list 54 papers and host 45. A paper presented at both conferences is kept once, giving 372 unique papers with a conference-hosted PDF. We drop 9 files that are not papers (poster images or two-page abstracts of fewer than five pages), leaving 363 unique papers. \autoref{tab:sample} summarizes the construction. Papers are identified by a hash of the normalized title; when a paper appears both in a regular session and in the poster session we keep the regular-session file.

Following \citet{dai2023}, we convert each PDF, finance and accounting, to text by extracting its text layer, and we use \ac{ocr} only when a file has no text layer, which is the case for one paper. Following \citet{hansen2018}, we remove running page headers, page footers, and page numbers; we retain footnotes, tables, references, and appendices because a referee reads them and because the longest paper (82{,}456 tokens) fits within the model's 128{,}000-token context window. The removal is restricted to the first and last three lines of each page and to lines that recur on at least 40\% of pages, so that numeric lines in the body, such as equation numbers and table cells, are never touched; the first occurrence of a recurring line is kept so that repeated table headers survive. The cleaned texts have a median length of 32{,}559 tokens and retain 99.8--99.9\% of the characters of the raw text layer.

\begin{table}[H]
    \centering
    \small
    \caption{\textbf{Sample Construction}\\This table reports the number of papers on the programs of the 2024 AFA Annual Meeting and the six 2024 NBER finance meetings (the Asset Pricing and Corporate Finance program meetings in spring and fall and the two Summer Institute meetings), the number for which the conference hosts the submitted PDF, and the number of usable files. A file is usable if it has at least five pages; the excluded files are poster images and two-page abstracts. Papers presented at more than one meeting are counted once in the last row; a paper appearing both in a regular session and in the poster session is assigned to the regular session.}
    \begingroup\renewcommand{\baselinestretch}{1}\selectfont
    \begin{tabular}{lccc}
\toprule
 & \shortstack{Papers on\\program} & \shortstack{Conference-\\hosted PDF} & \shortstack{Usable\\($\geq 5$ pages)} \\
\midrule
AFA 2024 Annual Meeting, regular sessions & 243 & 242 & 240 \\
AFA 2024 Annual Meeting, PhD poster session & 91 & 91 & 83 \\
NBER 2024 finance meetings (six meetings) & 54 & 45 & 45 \\
\midrule
Unique papers & 383 & 372 & 363 \\
\bottomrule
\end{tabular}

    \endgroup
    \label{tab:sample}
\end{table}

\paragraph{Accounting papers.}
\label{sec:appendix_acct}
The Conference on Financial Economics and Accounting hosts the full text of the papers on its program for its 2023 edition (Rutgers Business School, 3--4 November 2023) and its 2024 edition (University at Buffalo, 24--26 October 2024). We take the 24 papers in the accounting sessions of 2024 and the 21 papers in the accounting and joint sessions of 2023; four of the 2023 files are one-page abstracts, which leaves 41 papers. Their text is extracted and cleaned in the same way, and the prompt is the naive prompt with ``an accounting journal'' in place of ``a finance journal''.

\section{The Referee Protocol and Prompts}
\label{sec:prompt}

The knowledge cutoff of \texttt{gpt-4o-2024-11-20} (\autoref{sec:referee}) is also six months before the earliest \acs{nber} meeting in the sample; Section~\ref{sec:appendix_lookahead} tests whether the model knows the outcomes. Scoring each paper in a fresh session makes the scores absolute judgements rather than within-session rankings \citep{newton2026}. The prompt asks for the three-way recommendation that finance journals ask referees to make \citep{berk2017}. The response is constrained through the structured-output feature of the \ac{api} to a fixed schema with the fields \texttt{report} (string), \texttt{recommendation} (one of \texttt{Accept}, \texttt{Revise and resubmit}, \texttt{Reject}), \texttt{originality}, \texttt{correctness}, \texttt{significance}, \texttt{exposition} (integers in $[0,10]$), and \texttt{overall} (integer in $[0,100]$); the schema fixes the shape of the answer and is not visible to the model as an instruction.

The sampling temperature is left at the model's default, because a referee who uses a chat interface does not set it. Each paper is scored four times, because repeated evaluation at the default temperature is the standard way to measure the reliability of \acs{llm} judgements \citep{newton2026, ouyang2025, bini2026}. All 1{,}452 requests were submitted through the batch \acs{api}; 1{,}451 returned a parseable response.\footnote{In one response the model placed an incomplete report in the refusal field of the \acs{api}; that paper's scores are the mean of three draws.}

The prompt below is sent as a single user message, followed by the full text of the paper. The definitions of the four dimensions and the two scales are taken from Section~2.2 of \citet{newton2026}; the statements there are rephrased as questions.

\begin{quote}
\small
Please review this paper as a referee for a finance journal. Write a referee report and tell me whether it should be accepted, revised, or rejected. Also rate the paper from 0 to 10 on each of the following, and give an overall score from 0 to 100:

\begin{itemize}[nosep]
    \item Originality: does it present a genuinely new idea, angle, dataset, design, or synthesis relative to established academic knowledge?
    \item Correctness: are the analysis, empirical design, theoretical logic, and interpretation scientifically convincing?
    \item Significance: how important is the question, and how large is the contribution if the results are correct?
    \item Exposition: how clear, well organized, and well communicated is the paper?
\end{itemize}

\{paper\_text\}
\end{quote}

The guidance prompt of \autoref{sec:results_guidance} changes only the text before the rating instructions. It places the following sentence and the full text of \citet{berk2017} before the naive prompt:

\begin{quote}
\small
Before reviewing the paper below, please read the following article on how to referee research papers in finance and economics.

\{guide\_text\}
\end{quote}

The look-ahead probe of Section~\ref{sec:appendix_lookahead} is sent in a fresh session at temperature 0 with the same model. The response is constrained to the fields \texttt{status} (one of \texttt{Published}, \texttt{Accepted, forthcoming}, \texttt{Revise and resubmit}, \texttt{Submitted, under review}, \texttt{Not submitted or not published}, \texttt{Do not know}), \texttt{journal}, \texttt{year}, and \texttt{confidence} (integer in $[0,100]$). The line in brackets is included only in columns~(2) and (4) of \autoref{tab:probes_desc}.

\begin{quote}
\small
What is the publication status of the following paper? Tell me whether it has been published in an academic journal, accepted for publication, given a revise-and-resubmit, or submitted to a journal, and if so which journal and in which year. If you do not know, say so.

Title: \{title\}\\
{[}Authors: \{authors\}{]}
\end{quote}

\section{Publication Outcomes}
\label{sec:appendix_outcomes}

A calendar cutoff would give papers presented in November 2024 ten months less than papers presented in January, which is why we measure publication within a fixed window of two years after each paper's meeting (\autoref{sec:data}).

We follow the two-stage procedure of \citet{hou2025}, who track the publication status of 6{,}805 finance conference papers. In the automated stage we search for each paper's title in Crossref, OpenAlex, Semantic Scholar, and Google Scholar and match a candidate journal article when the normalized-title similarity is at least 0.85 and at least one author surname is shared; matches with title similarity between 0.60 and 0.85 are flagged for manual review. Records whose venue is a preprint server, a working-paper series, or a web domain are never treated as journal articles, and the date of a matched article is taken from Crossref because aggregators that merge a working paper with its published version sometimes report the working paper's date. In the manual stage, which \citet{hou2025} describe as necessary because ``changes in paper titles'' are ``the most frequent obstacle when tracking papers,'' we search Crossref by the authors' surnames for every paper still unmatched, compare the abstracts of candidate articles whose author lists coincide with the program entry, and read the acknowledgement footnotes of the published versions. This stage recovers nine published papers whose titles changed before publication, eight of which state their conference title in a footnote and eight of which thank the 2024 \acs{afa} meeting. Forty-two grey-zone cases were decided by hand. \citet{hou2025} report that title-based matching alone misses about two thirds of eventual publications over a 6--10 year horizon; over our shorter horizon it misses 11\%.

The outcome variables follow the finance journal groups of \citet{dai2023}; \citet{bajo2020} obtain the same top group from promotion data and \citet{currie2020} the same second tier from scholar assessments. Besides ${\it Top3Fin}_i$, their main outcome, ${\it Top5Fin}_i$ adds the \emph{Journal of Financial and Quantitative Analysis} and the \emph{Review of Finance}, and ${\it Top10Fin}_i$ further adds the \emph{Journal of Corporate Finance}, the \emph{Journal of Banking \& Finance}, \emph{Financial Management}, the \emph{Journal of Financial Markets}, and the \emph{Journal of Financial Intermediation} (\citealp{dai2023}, Internet Appendix). Because the sample is drawn from finance conferences, we use only these finance outcomes. \autoref{tab:outcomes} reports the distribution for the 276 papers of \autoref{tab:logit_main}: within two years of the meeting, 18 (6.5\%) are in a top-three finance journal, 20 (7.2\%) in a top-five finance journal, and 22 (8.0\%) in a top-ten finance journal. Among all 363 papers, including the poster session, the counts are 21 (5.8\%), 24 (6.6\%), and 27 (7.4\%). The window is short by the standards of \citet{hou2025}, who find that many conference papers take more than five years to be accepted, so the unpublished group contains papers that will eventually be published, which biases our estimates toward zero.

\begin{table}[H]
    \centering
    \small
    \caption{\textbf{Publication Outcomes}\\This table reports the publication status, two years after their meeting, of the 276 papers presented at the 2024 AFA Annual Meeting and the six 2024 NBER finance meetings outside the AFA PhD student poster session that enter the regressions (Panel A) and of the 41 accounting papers presented at the 2023 and 2024 Conferences on Financial Economics and Accounting (Panel B). The outcomes were collected on 18 September 2026 for the finance papers and on 5 October 2026 for the accounting papers. A paper is counted as published in a journal when a matching article is online within two years of its meeting; matching follows the two-stage procedure of \citet{hou2025} described in Section~\ref{sec:appendix_outcomes}. The finance outcomes follow \citet{dai2023}: the top three finance journals (the \emph{Journal of Finance}, the \emph{Journal of Financial Economics}, and the \emph{Review of Financial Studies}), the top five (adding the \emph{Journal of Financial and Quantitative Analysis} and the \emph{Review of Finance}), and the top ten (adding the \emph{Journal of Corporate Finance}, the \emph{Journal of Banking \& Finance}, \emph{Financial Management}, the \emph{Journal of Financial Markets}, and the \emph{Journal of Financial Intermediation}). The top three accounting journals are \emph{The Accounting Review}, the \emph{Journal of Accounting Research}, and the \emph{Journal of Accounting and Economics}; three accounting papers are published in journals outside these three.}
    \begingroup\renewcommand{\baselinestretch}{1}\selectfont
    \textit{Panel A: Finance}\\[2pt]
    \begin{tabular}{lcc}
\toprule
 & Papers & Share (\%) \\
\midrule
${\it Top3Fin}_i = 1$ (JF, JFE, RFS) & 18 & 6.5 \\
${\it Top5Fin}_i = 1$ (top three plus JFQA, RoF) & 20 & 7.2 \\
${\it Top10Fin}_i = 1$ (top five plus JCF, JBF, FM, JFM, JFI) & 22 & 8.0 \\
Journal outside the top ten finance journals & 5 & 1.8 \\
Not published & 249 & 90.2 \\
\midrule
Papers & 276 & 100.0 \\
\bottomrule
\end{tabular}

    \vspace{8pt}
    \textit{Panel B: Accounting}\\[2pt]
    \begin{tabular}{lcc}
\toprule
 & Papers & Share (\%) \\
\midrule
${\it Top3Acc}_i = 1$ (TAR, JAR, JAE) & 5 & 12.2 \\
Journal outside the top three accounting journals & 3 & 7.3 \\
Not published & 33 & 80.5 \\
\midrule
Papers & 41 & 100.0 \\
\bottomrule
\end{tabular}

    \endgroup
    \label{tab:outcomes}
\end{table}

\section{Author Reputation}
\label{sec:appendix_reputation}

This section describes the reputation measures that enter the controls of equation~\eqref{eq:logit} and the interaction specification of \autoref{tab:logit_reputation}, which follows Table~11 of \citet{dai2023} and asks whether the score is more informative for some author groups than for others.

\citet{dai2023} measure author standing by a top-twenty school indicator and a coauthor-network centrality computed over the whole SSRN finance network. We keep their school indicator: ${\it Top20School}_i$ equals one if any author's affiliation on the conference program is one of the twenty schools of the UT Dallas research ranking computed on publications in \acs{jf}, \acs{jfe}, and \acs{rfs}, using the worldwide ranking for 2019--2023, the last five-year window that ends before the 2024 meetings. The centrality is not reproducible on a conference sample, and we replace it with the publication record of the authors, in the spirit of the prolific-author measure of \citet{card2020}, and with their editorial experience (${\it Editorial}_i$). For each author we obtain the OpenAlex profile \citep{priem2022}, taken from the authorship list of the paper's own OpenAlex record where one exists (75\% of authors) and otherwise from an author search by name disambiguated by the program affiliation; 3\% of authors, mostly PhD students and authors with very common names, cannot be matched. ${\it Log\ Prior\ Top}_i$ is the log of one plus the largest number of \acs{jf}, \acs{jfe}, or \acs{rfs} articles published before 2024 by any author of the paper. It is computed on records dated before 2024, so it is the reputation an editor would have observed at the time of the conference. We restrict works to the economics and business fields because OpenAlex merges some scholars with namesakes in other fields; a profile with no economics or business publication is treated as unmatched.

The sort variables of the interaction specification follow Table~11 of \citet{dai2023}, adapted to our data: their sorts are the school indicator, an indicator for a top-5\% value of their centrality measure, and an early-stage indicator. Ours are ${\it Established}_i$, one if any author has a prior \acs{jf}, \acs{jfe}, or \acs{rfs} article, the indicator form of ${\it Log\ Prior\ Top}_i$ and the prolific-author margin of \citet{card2020}; ${\it Editorial}_i$ (\autoref{sec:data}), which marks the authors closest to the editorial process \citep{brogaard2014}; and ${\it Early\ Stage}_i$, one if every matched author's first publication in OpenAlex is within the ten years before the conference, the ``early stage researcher'' definition of \citet{dai2023}. A sort variable enters the specification once, so when it is itself a control it replaces it.

\paragraph{Results.}
\label{sec:appendix_reputation_results}
\autoref{tab:logit_reputation} reports the interactions. Of the 276 papers of \autoref{tab:logit_main}, 11\% are written entirely by early-stage researchers; the shares with an established author and with an editor among the authors are in \autoref{tab:sum_scores}.

Column~(1) sorts on ${\it Established}_i$. For papers whose authors have no prior top-three article the score has a coefficient of 0.201 ($z = 1.13$), and the interaction is positive (0.337, $z = 1.11$), an implied coefficient of about 0.54 for papers with an established author; the difference is not significant. Column~(2) sorts on ${\it Editorial}_i$. For papers without an editor or associate editor among the authors the score has a coefficient of 0.693 ($z = 2.97$), and the interaction is $-0.599$ ($z = -1.34$), an implied coefficient of about 0.09 for papers by editors. Column~(3) sorts on ${\it Early\ Stage}_i$. The score is a strong predictor for teams with at least one experienced author (0.516, $z = 2.33$) and weaker for all-early-stage teams; the interaction is $-0.350$ ($z = -1.13$). With 30 early-stage papers the estimate is imprecise, but it points in the direction of the model being least informative where a referee's judgement would be most valuable, for authors without a track record. The same specification on the 236 papers whose authors' OpenAlex profiles all come from the paper's own OpenAlex record, excluding the 40 papers with at least one profile matched by a name search, gives a coefficient of 0.467 ($z = 2.28$) on the score for ${\it Top3Fin}_i$ (untabulated).

\begin{table}[H]
    \centering
    \small
    \caption{\textbf{The Predictability of Publication by Author Standing}\\This table presents logit estimates of equation~\eqref{eq:logit} in which the standardized overall score is interacted with a sort variable, following Table~11 of \citet{dai2023}. Column (1) sorts on ${\it Established}_i$, one if any author has an article published before 2024 in the \emph{Journal of Finance}, the \emph{Journal of Financial Economics}, or the \emph{Review of Financial Studies}; column (2) on ${\it Editorial}_i$, one if any author served as an editor or associate editor of these journals in 2014--2024; and column (3) on ${\it Early\ Stage}_i$, one if every matched author's first publication is within the ten years before the conference (Section~\ref{sec:appendix_reputation}). The sample is the 276 papers presented at the 2024 AFA Annual Meeting and the six 2024 NBER finance meetings outside the AFA PhD student poster session that enter the regressions (\autoref{sec:spec}). The outcome is ${\it Top3Fin}_i$, one if the paper is online at the \emph{Journal of Finance}, the \emph{Journal of Financial Economics}, or the \emph{Review of Financial Studies} within two years of its meeting. The overall score is the mean over four independent draws of GPT-4o (\texttt{gpt-4o-2024-11-20}) under the naive prompt reproduced in Section~\ref{sec:prompt}, standardized across papers. The coefficient on ${\it Overall}_i$ is the effect for papers with the sort variable equal to zero; the interaction is the difference for papers with the sort variable equal to one. All specifications include the log length of the paper text, the number of authors, ${\it Top20School}_i$, ${\it Log\ Prior\ Top}_i$, ${\it Editorial}_i$ (which in column (2) enters only as the sort variable), and a conference (AFA versus NBER) fixed effect. All variables are defined in \autoref{tab:var_des}. Robust $z$-statistics in parentheses. ***, **, and * denote significance at the 1\%, 5\%, and 10\% levels.}
    \begingroup\renewcommand{\baselinestretch}{1}\selectfont
    {
\def\sym#1{\ifmmode^{#1}\else\(^{#1}\)\fi}
\begin{tabular}{l*{3}{c}}
\toprule
                    &\multicolumn{3}{c}{${\it Top3Fin}_i$}\\\cmidrule(lr){2-4}
                    &\multicolumn{1}{c}{(1)}&\multicolumn{1}{c}{(2)}&\multicolumn{1}{c}{(3)}\\
Sort variable       &\multicolumn{1}{c}{${\it Established}_i$}&\multicolumn{1}{c}{${\it Editorial}_i$}&\multicolumn{1}{c}{${\it Early\ Stage}_i$}\\
\midrule
${\it Overall}_i$& 0.201  &  0.693\sym{***}  &  0.516\sym{**} \\
                    & (1.13)  &  (2.97)  &  (2.33) \\
${\it Sort\ Variable}_i$& 1.062  &  0.133  &  0.015 \\
                    & (0.76)  &  (0.24)  &  (0.01) \\
${\it Overall}_i \times {\it Sort\ Variable}_i$& 0.337  &  -0.599  &  -0.350 \\
                    & (1.11)  &  (-1.34)  &  (-1.13) \\
${\it Log\ Length}_i$& 0.033  &  0.070  &  0.025 \\
                    & (0.06)  &  (0.12)  &  (0.04) \\
${\it N\ Authors}_i$& 0.573\sym{**}  &  0.601\sym{**}  &  0.594\sym{**} \\
                    & (2.09)  &  (2.37)  &  (2.31) \\
${\it Top20School}_i$& 0.153  &  0.152  &  0.169 \\
                    & (0.31)  &  (0.30)  &  (0.33) \\
${\it Log\ Prior\ Top}_i$& -0.107  &  0.112  &  0.100 \\
                    & (-0.26)  &  (0.41)  &  (0.29) \\
${\it Editorial}_i$& 0.005  &    &  -0.054 \\
                    & (0.01)  &    &  (-0.09) \\
\midrule
Conference FE       & Y  &  Y  &  Y \\
N Obs               & 276  &  276  &  276 \\
Pseudo R$^2$        & 0.080  &  0.079  &  0.072 \\
\bottomrule
\end{tabular}
}

    \endgroup
    \label{tab:logit_reputation}
\end{table}

\section{Look-Ahead Bias}
\label{sec:appendix_lookahead}

Because every publication outcome in the sample occurs after the model's knowledge cutoff (\autoref{sec:results_memory}), the outcome itself cannot be in the training data, the design that \citet{lopezlira2026} use to rule out look-ahead bias in return prediction; \citet{sarkar2024lookahead} and \citet{levy2026caution} document how much such bias can distort \acs{llm}-based results in finance and accounting. A signal of the outcome can be. An author's website or CV may list a paper as a revise-and-resubmit or as forthcoming at a journal before the cutoff, and a model that has read it could recognize the paper and score it accordingly.

We test for this channel by asking the model directly, as \citet{lopezlira2025} do for economic data the model may have memorized. In a fresh session we ask the scoring model for the publication status of each paper: whether it is published, accepted, under revise-and-resubmit, or submitted, at which journal, and in which year, with the option to answer that it does not know (Section~\ref{sec:prompt}). We vary the identifying information in the spirit of \citet{glasserman2024}, who vary the firm identifiers in news text: we give the model the conference title alone, and then the title and the authors, the pair through which a CV entry would link the paper to a journal. We set the temperature to 0, under which a memorized answer is most likely to surface. We ask about the unpublished papers as well as the published ones, because the unpublished papers show how often the model claims a publication that did not happen. For each information set we define ${\it Claim}_i$ as one if the model answers that the paper is published or accepted. Following \citet{gao2025}, we add ${\it Claim}_i$ and its interaction with the score to \eqref{eq:logit}, one information set at a time. If the score predicts publication because the model knows the outcome, its coefficient should fall once ${\it Claim}_i$ is controlled for, and the interaction should be positive.

\autoref{sec:results_memory} reports the model's main answers (columns~(1)--(2) of \autoref{tab:probes_desc}). In addition, with the title and authors the model calls 10.1\% of the papers published. These claims are guesses. The model makes them for 10.4\% of the papers that remain unpublished at the end of the outcome window and for 7.4\% of those that are published, and the claimed years are 2023 or earlier, whereas the earliest publication in the sample appears in late October 2023.

Of the 27 papers published in a journal within the outcome window, the model names the correct journal for one given the title and for another given the title and authors. We read the submission and acceptance dates from the published articles: both were submitted to the journal after the cutoff, so no editorial decision on them could have been public. Asked about every conference paper, the model also names the journal of one paper that was submitted in May 2023 and accepted in February 2024, so a revise-and-resubmit could have been public; we exclude it from every regression. For the 41 accounting papers (columns~(3)--(4) of \autoref{tab:probes_desc}), the model names the journal of none of the eight papers published within the window, so no accounting paper is excluded.

The model's claims carry no information about publication (\autoref{tab:logit_lookahead}). With the claim and its interaction added, the coefficient on the overall score for ${\it Top3Fin}_i$ stays close to the 0.495 of \autoref{tab:logit_main} for both information sets (\autoref{sec:results_memory}). Neither the claim nor the interaction is significant in any column; for the title and authors the interaction is 0.291 ($z = 0.69$). For ${\it Top5Fin}_i$ and ${\it Top10Fin}_i$ the coefficient on the overall score is between 0.435 and 0.578 and significant at the 5\% level in every column. On the 248 papers the model does not call published given the title and authors, the coefficient on the overall score is 0.438 ($z = 2.24$) for ${\it Top3Fin}_i$ and 0.522 ($z = 2.74$) for ${\it Top10Fin}_i$ (untabulated). The score is somewhat higher for papers the model calls published (0.41 standard deviations, $t = 2.03$, untabulated), but this component does not predict which papers are published.

\begin{table}[H]
    \centering
    \small
    \caption{\textbf{What Does the Model Know About Publication Outcomes?}\\This table reports what GPT-4o (\texttt{gpt-4o-2024-11-20}), the model whose scores the paper studies, answers when asked in a fresh session at temperature 0 for the publication status of each paper (Section~\ref{sec:appendix_lookahead}); the question is reproduced in Section~\ref{sec:prompt}. The model receives the conference title (columns 1 and 3) or the title and authors (columns 2 and 4), and gives one answer per paper. Columns (1)--(2) report the 276 papers presented at the 2024 AFA Annual Meeting and the six 2024 NBER finance meetings outside the AFA PhD student poster session that enter the regressions, and columns (3)--(4) the 41 accounting papers presented at the 2023 and 2024 Conferences on Financial Economics and Accounting. The top rows report the distribution of answers. The bottom rows count, among the papers published in a journal within two years of their meeting, the papers for which the model names the correct journal, and among these the papers submitted to that journal, or online, before the model's knowledge cutoff of 1 October 2023, as dated in the published article. A repeat of the finance title-and-authors probe with identical requests gives the same answer for 97.5\% of the papers.}
    \begingroup\renewcommand{\baselinestretch}{1}\selectfont\setlength{\tabcolsep}{4pt}
    \begin{tabular}{lcccc}
\toprule
 & \multicolumn{2}{c}{Finance} & \multicolumn{2}{c}{Accounting} \\
\cmidrule(lr){2-3}\cmidrule(lr){4-5}
 & Title & Title and authors & Title & Title and authors \\
 & (1) & (2) & (3) & (4) \\
\midrule
\multicolumn{5}{l}{\textit{Answer (share of the papers)}} \\
Papers & 276 & 276 & 41 & 41 \\
Do not know & 0.964 & 0.899 & 0.976 & 1.000 \\
Published or accepted & 0.036 & 0.101 & 0.024 & 0.000 \\
Revise and resubmit or under review & 0.000 & 0.000 & 0.000 & 0.000 \\
\midrule
\multicolumn{5}{l}{\textit{Papers published in a journal within the outcome window}} \\
Papers & 27 & 27 & 8 & 8 \\
Journal named correctly & 1 & 1 & 0 & 0 \\
\quad of which submitted or online before the cutoff & 0 & 0 & 0 & 0 \\
\bottomrule
\end{tabular}

    \endgroup
    \label{tab:probes_desc}
\end{table}

\begin{table}[H]
    \centering
    \small
    \caption{\textbf{LLM Scores and Publication, Controlling for What the Model Claims to Know}\\This table presents logit estimates of equation~\eqref{eq:logit} with ${\it Claim}_i$ and its interaction with the overall score added \citep{gao2025}. ${\it Claim}_i$ equals one if GPT-4o, asked in a fresh session at temperature 0 for the publication status of the paper, answers that it is published or accepted when given the conference title (odd columns) or the title and authors (even columns), as in \autoref{tab:probes_desc}. Columns (1)--(2) report ${\it Top3Fin}_i$ (online at the \emph{Journal of Finance}, the \emph{Journal of Financial Economics}, or the \emph{Review of Financial Studies}), columns (3)--(4) ${\it Top5Fin}_i$ (the top three finance journals plus the \emph{Journal of Financial and Quantitative Analysis} and the \emph{Review of Finance}), and columns (5)--(6) ${\it Top10Fin}_i$ (the top five plus the \emph{Journal of Corporate Finance}, the \emph{Journal of Banking \& Finance}, \emph{Financial Management}, the \emph{Journal of Financial Markets}, and the \emph{Journal of Financial Intermediation}), all measured within two years of each paper's meeting. The sample is the 276 papers presented at the 2024 AFA Annual Meeting and the six 2024 NBER finance meetings outside the AFA PhD student poster session that enter the regressions (\autoref{sec:spec}). The overall score is the mean over four independent draws of GPT-4o (\texttt{gpt-4o-2024-11-20}) under the naive prompt reproduced in Section~\ref{sec:prompt}, standardized across papers. All specifications include the log length of the paper text, the number of authors, ${\it Top20School}_i$, ${\it Log\ Prior\ Top}_i$, ${\it Editorial}_i$, and a conference (AFA versus NBER) fixed effect. All variables are defined in \autoref{tab:var_des}. Robust $z$-statistics in parentheses. ***, **, and * denote significance at the 1\%, 5\%, and 10\% levels.}
    \begingroup\renewcommand{\baselinestretch}{1}\selectfont
    {
\def\sym#1{\ifmmode^{#1}\else\(^{#1}\)\fi}
\begin{tabular}{l*{6}{c}}
\toprule
                    &\multicolumn{2}{c}{${\it Top3Fin}_i$}&\multicolumn{2}{c}{${\it Top5Fin}_i$}&\multicolumn{2}{c}{${\it Top10Fin}_i$}\\\cmidrule(lr){2-3}\cmidrule(lr){4-5}\cmidrule(lr){6-7}
                    &\multicolumn{1}{c}{(1)}&\multicolumn{1}{c}{(2)}&\multicolumn{1}{c}{(3)}&\multicolumn{1}{c}{(4)}&\multicolumn{1}{c}{(5)}&\multicolumn{1}{c}{(6)}\\
                    &\multicolumn{1}{c}{Title}&\multicolumn{1}{c}{\shortstack{Title and\\authors}}&\multicolumn{1}{c}{Title}&\multicolumn{1}{c}{\shortstack{Title and\\authors}}&\multicolumn{1}{c}{Title}&\multicolumn{1}{c}{\shortstack{Title and\\authors}}\\
\midrule
${\it Overall}_i$& 0.491\sym{**}  &  0.466\sym{**}  &  0.456\sym{**}  &  0.435\sym{**}  &  0.578\sym{***}  &  0.568\sym{***} \\
                    & (2.29)  &  (2.18)  &  (2.34)  &  (2.25)  &  (2.88)  &  (2.80) \\
${\it Claim}_i$& -0.108  &  -0.313  &  -0.448  &  -0.505  &  -0.469  &  -0.525 \\
                    & (-0.13)  &  (-0.43)  &  (-0.53)  &  (-0.73)  &  (-0.55)  &  (-0.77) \\
${\it Overall}_i \times {\it Claim}_i$& 0.113  &  0.291  &  0.125  &  0.288  &  -0.072  &  0.151 \\
                    & (0.20)  &  (0.69)  &  (0.23)  &  (0.71)  &  (-0.13)  &  (0.38) \\
${\it Log\ Length}_i$& 0.019  &  0.025  &  0.143  &  0.163  &  -0.094  &  -0.064 \\
                    & (0.03)  &  (0.04)  &  (0.26)  &  (0.30)  &  (-0.17)  &  (-0.11) \\
${\it N\ Authors}_i$& 0.595\sym{**}  &  0.593\sym{**}  &  0.681\sym{***}  &  0.656\sym{***}  &  0.681\sym{***}  &  0.642\sym{***} \\
                    & (2.44)  &  (2.35)  &  (2.88)  &  (2.77)  &  (2.89)  &  (2.73) \\
${\it Top20School}_i$& 0.171  &  0.177  &  -0.207  &  -0.198  &  -0.392  &  -0.385 \\
                    & (0.34)  &  (0.35)  &  (-0.41)  &  (-0.39)  &  (-0.82)  &  (-0.81) \\
${\it Log\ Prior\ Top}_i$& 0.110  &  0.117  &  0.093  &  0.112  &  0.035  &  0.063 \\
                    & (0.39)  &  (0.42)  &  (0.34)  &  (0.42)  &  (0.13)  &  (0.24) \\
${\it Editorial}_i$& -0.060  &  -0.070  &  0.396  &  0.379  &  0.310  &  0.295 \\
                    & (-0.10)  &  (-0.12)  &  (0.70)  &  (0.66)  &  (0.56)  &  (0.53) \\
\midrule
Conference FE       & Y  &  Y  &  Y  &  Y  &  Y  &  Y \\
N Obs               & 276  &  276  &  276  &  276  &  276  &  276 \\
Pseudo R$^2$        & 0.072  &  0.073  &  0.087  &  0.088  &  0.093  &  0.093 \\
\bottomrule
\end{tabular}
}

    \endgroup
    \label{tab:logit_lookahead}
\end{table}

\section{Additional Results}
\label{sec:appendix_additional}

\paragraph{Specification details.} \citet{dai2023} estimate the publication outcome by logit with year and field fixed effects and cluster standard errors by year; our sample is a single conference year, so the fixed effect distinguishes \acs{afa} from \acs{nber} papers and we report heteroskedasticity-robust standard errors. The length and author-count controls correspond to the log page count and coauthor count of \citet{dai2023}, and the prior top-three record and ${\it Editorial}_i$ take the place of their coauthor-network centrality. The dimension scores enter one at a time because the overall score is, by the design of the protocol, a summary of the four dimensions.

\paragraph{Controls.} Among the controls in \autoref{tab:logit_main}, only the number of authors is significant throughout: each additional author raises the log odds of top-three publication by about 0.60 ($z = 2.45$ in column~1); the other controls are insignificant (\autoref{sec:results_main}). The coauthor effect contrasts with the negative coauthor coefficient that \citet{dai2023} find in the universe of SSRN postings; the difference is consistent with conference selection having already removed the papers for which team size proxies for lower quality.

\paragraph{Broader journal sets.} \autoref{tab:logit_fin_broad} reports the estimates for ${\it Top5Fin}_i$ and ${\it Top10Fin}_i$, which Sections~\ref{sec:results_main} and \ref{sec:results_dimensions} summarize. Exposition predicts both outcomes (0.636, $z = 2.66$, and 0.633, $z = 2.76$), correctness does so at the 10\% level (0.546, $z = 1.74$, and 0.551, $z = 1.88$), and originality and significance do not.

\begin{table}[H]
    \centering
    \small
    \caption{\textbf{The LLM Referee's Scores and Publication in the Broader Finance Journal Sets}\\This table presents logit estimates of equation~\eqref{eq:logit} for the two broader finance outcomes of \citet{dai2023}, measured within two years of each paper's meeting: ${\it Top5Fin}_i$ (the top three finance journals plus the \emph{Journal of Financial and Quantitative Analysis} and the \emph{Review of Finance}) in columns (1)--(5), and ${\it Top10Fin}_i$ (the top five plus the \emph{Journal of Corporate Finance}, the \emph{Journal of Banking \& Finance}, \emph{Financial Management}, the \emph{Journal of Financial Markets}, and the \emph{Journal of Financial Intermediation}) in columns (6)--(10). The sample is the 276 papers presented at the 2024 AFA Annual Meeting and the six 2024 NBER finance meetings outside the AFA PhD student poster session that enter the regressions (\autoref{sec:spec}). Scores are the means over four independent draws of GPT-4o (\texttt{gpt-4o-2024-11-20}) under the naive prompt reproduced in Section~\ref{sec:prompt}, standardized across papers; each column enters one score, and a coefficient is the change in log odds per standard deviation of the score. All specifications include the log length of the paper text, the number of authors, ${\it Top20School}_i$, ${\it Log\ Prior\ Top}_i$, ${\it Editorial}_i$, and a conference (AFA versus NBER) fixed effect. All variables are defined in \autoref{tab:var_des}. Robust $z$-statistics in parentheses. ***, **, and * denote significance at the 1\%, 5\%, and 10\% levels.}
    \begingroup\renewcommand{\baselinestretch}{1}\selectfont
    \resizebox{\linewidth}{!}{{
\def\sym#1{\ifmmode^{#1}\else\(^{#1}\)\fi}
\begin{tabular}{l*{10}{c}}
\toprule
                    &\multicolumn{5}{c}{${\it Top5Fin}_i$}                                                                        &\multicolumn{5}{c}{${\it Top10Fin}_i$}                                                                       \\\cmidrule(lr){2-6}\cmidrule(lr){7-11}
                    &\multicolumn{1}{c}{(1)}         &\multicolumn{1}{c}{(2)}         &\multicolumn{1}{c}{(3)}         &\multicolumn{1}{c}{(4)}         &\multicolumn{1}{c}{(5)}         &\multicolumn{1}{c}{(6)}         &\multicolumn{1}{c}{(7)}         &\multicolumn{1}{c}{(8)}         &\multicolumn{1}{c}{(9)}         &\multicolumn{1}{c}{(10)}         \\
\midrule
${\it Overall}_i$   &       0.451\sym{**} &                     &                     &                     &                     &       0.561\sym{***}&                     &                     &                     &                     \\
                    &      (2.39)         &                     &                     &                     &                     &      (2.92)         &                     &                     &                     &                     \\
${\it Originality}_i$&                     &      -0.115         &                     &                     &                     &                     &      -0.115         &                     &                     &                     \\
                    &                     &     (-0.56)         &                     &                     &                     &                     &     (-0.62)         &                     &                     &                     \\
${\it Significance}_i$&                     &                     &      -0.008         &                     &                     &                     &                     &       0.019         &                     &                     \\
                    &                     &                     &     (-0.03)         &                     &                     &                     &                     &      (0.08)         &                     &                     \\
${\it Correctness}_i$&                     &                     &                     &       0.546\sym{*}  &                     &                     &                     &                     &       0.551\sym{*}  &                     \\
                    &                     &                     &                     &      (1.74)         &                     &                     &                     &                     &      (1.88)         &                     \\
${\it Exposition}_i$&                     &                     &                     &                     &       0.636\sym{***}&                     &                     &                     &                     &       0.633\sym{***}\\
                    &                     &                     &                     &                     &      (2.66)         &                     &                     &                     &                     &      (2.76)         \\
${\it Log\ Length}_i$&       0.145         &       0.180         &       0.178         &       0.102         &      -0.179         &      -0.083         &      -0.038         &      -0.051         &      -0.137         &      -0.403         \\
                    &      (0.27)         &      (0.35)         &      (0.33)         &      (0.18)         &     (-0.32)         &     (-0.15)         &     (-0.07)         &     (-0.09)         &     (-0.25)         &     (-0.72)         \\
${\it N\ Authors}_i$&       0.664\sym{***}&       0.625\sym{***}&       0.626\sym{***}&       0.668\sym{***}&       0.689\sym{***}&       0.650\sym{***}&       0.595\sym{***}&       0.595\sym{***}&       0.642\sym{***}&       0.664\sym{***}\\
                    &      (2.90)         &      (2.72)         &      (2.74)         &      (2.89)         &      (3.01)         &      (2.89)         &      (2.65)         &      (2.67)         &      (2.86)         &      (3.02)         \\
${\it Top20School}_i$&      -0.203         &      -0.105         &      -0.134         &      -0.164         &      -0.221         &      -0.385         &      -0.278         &      -0.310         &      -0.343         &      -0.402         \\
                    &     (-0.40)         &     (-0.21)         &     (-0.27)         &     (-0.33)         &     (-0.46)         &     (-0.81)         &     (-0.59)         &     (-0.67)         &     (-0.73)         &     (-0.89)         \\
${\it Log\ Prior\ Top}_i$&       0.100         &       0.078         &       0.088         &       0.114         &       0.100         &       0.051         &       0.024         &       0.032         &       0.058         &       0.036         \\
                    &      (0.38)         &      (0.29)         &      (0.33)         &      (0.45)         &      (0.35)         &      (0.20)         &      (0.09)         &      (0.12)         &      (0.23)         &      (0.13)         \\
${\it Editorial}_i$ &       0.372         &       0.460         &       0.460         &       0.325         &       0.399         &       0.278         &       0.394         &       0.394         &       0.260         &       0.325         \\
                    &      (0.66)         &      (0.83)         &      (0.83)         &      (0.56)         &      (0.71)         &      (0.50)         &      (0.73)         &      (0.73)         &      (0.47)         &      (0.59)         \\
\midrule
Conference FE       &           Y         &           Y         &           Y         &           Y         &           Y         &           Y         &           Y         &           Y         &           Y         &           Y         \\
N Obs               &         276         &         276         &         276         &         276         &         276         &         276         &         276         &         276         &         276         &         276         \\
Pseudo R$^2$        &       0.086         &       0.067         &       0.066         &       0.095         &       0.110         &       0.091         &       0.060         &       0.059         &       0.090         &       0.103         \\
\bottomrule
\end{tabular}
}
}
    \endgroup
    \label{tab:logit_fin_broad}
\end{table}

\paragraph{GPT-4o and GPT-5 scores.} \autoref{tab:gpt5_change} compares the scores that the two models give the same papers (\autoref{sec:results_models}). GPT-5 scores correctness lower for every paper, by 1.74 points for the finance papers and 2.28 points for the accounting papers, and scores originality and significance lower on average; its overall score is unchanged for the finance papers and 3.46 points lower for the accounting papers.

\begin{table}[H]
    \centering
    \small
    \caption{\textbf{GPT-4o and GPT-5 Scores of the Same Papers}\\This table compares the scores that GPT-4o (\texttt{gpt-4o-2024-11-20}) and GPT-5 (\texttt{gpt-5-2025-08-07}, minimal reasoning effort) give the same papers under the naive prompt. Panel A reports the 276 papers presented at the 2024 AFA Annual Meeting and the six 2024 NBER finance meetings outside the AFA PhD student poster session that enter the regressions, and Panel B the 41 accounting papers presented at the 2023 and 2024 Conferences on Financial Economics and Accounting. Each paper's score is the mean over its four draws. The difference is the mean of GPT-5's score minus GPT-4o's score, with the paired $t$-statistic in parentheses; ***, **, and * denote significance at the 1\%, 5\%, and 10\% levels. Higher, Lower, and Same report the percentage of papers for which GPT-5's score is above, below, or equal to GPT-4o's.}
    \begingroup\renewcommand{\baselinestretch}{1}\selectfont\setlength{\tabcolsep}{3pt}
    {
\def\sym#1{\ifmmode^{#1}\else\(^{#1}\)\fi}
\begin{tabular}{l*{7}{c}}
\toprule
 & \multicolumn{2}{c}{Mean} & \multicolumn{2}{c}{Difference} & \multicolumn{3}{c}{GPT-5 score (\% of papers)} \\
\cmidrule(lr){2-3}\cmidrule(lr){4-5}\cmidrule(lr){6-8}
 & GPT-4o & GPT-5 & GPT-5 $-$ 4o & $t$ & Higher & Lower & Same \\
\midrule
\multicolumn{8}{l}{\textit{Panel A: Finance (276 papers)}} \\
${\it Overall}_i$ & 81.92 & 81.93 & 0.01 & (0.03) & 57.6 & 40.6 & 1.8 \\
${\it Originality}_i$ & 8.31 & 8.02 & -0.29\sym{***} & (-7.38) & 26.4 & 58.3 & 15.2 \\
${\it Significance}_i$ & 8.12 & 7.89 & -0.22\sym{***} & (-5.84) & 29.7 & 50.0 & 20.3 \\
${\it Correctness}_i$ & 8.40 & 6.66 & -1.74\sym{***} & (-43.77) & 0.0 & 100.0 & 0.0 \\
${\it Exposition}_i$ & 7.21 & 7.18 & -0.02 & (-0.67) & 40.2 & 39.5 & 20.3 \\
\midrule
\multicolumn{8}{l}{\textit{Panel B: Accounting (41 papers)}} \\
${\it Overall}_i$ & 82.54 & 79.08 & -3.46\sym{***} & (-5.96) & 19.5 & 78.0 & 2.4 \\
${\it Originality}_i$ & 8.18 & 7.64 & -0.54\sym{***} & (-7.09) & 9.8 & 82.9 & 7.3 \\
${\it Significance}_i$ & 8.06 & 7.45 & -0.61\sym{***} & (-7.66) & 9.8 & 87.8 & 2.4 \\
${\it Correctness}_i$ & 8.51 & 6.23 & -2.28\sym{***} & (-27.50) & 0.0 & 100.0 & 0.0 \\
${\it Exposition}_i$ & 7.23 & 7.03 & -0.20\sym{***} & (-3.35) & 12.2 & 56.1 & 31.7 \\
\bottomrule
\end{tabular}
}

    \endgroup
    \label{tab:gpt5_change}
\end{table}

\paragraph{Correlations of the scores.} \autoref{tab:dim_corr} reports the correlations across papers of the five scores that GPT-4o gives under the naive prompt, which \autoref{sec:results_dimensions} discusses.

\begin{table}[H]
    \centering
    \small
    \caption{\textbf{Correlations of the LLM Referee's Scores}\\This table reports Pearson correlations across papers of the scores that GPT-4o (\texttt{gpt-4o-2024-11-20}) gives under the naive prompt. Panel A reports the 276 papers presented at the 2024 AFA Annual Meeting and the six 2024 NBER finance meetings outside the AFA PhD student poster session that enter the regressions, and Panel B the 41 accounting papers presented at the 2023 and 2024 Conferences on Financial Economics and Accounting. Each paper's score is the mean over its four draws.}
    \begingroup\renewcommand{\baselinestretch}{1}\selectfont
    {
\def\sym#1{\ifmmode^{#1}\else\(^{#1}\)\fi}
\begin{tabular}{l*{5}{c}}
\toprule
 & (1) & (2) & (3) & (4) & (5) \\
\midrule
\multicolumn{6}{l}{\textit{Panel A: Finance (276 papers)}} \\
(1) ${\it Overall}_i$ & 1 &  &  &  &  \\
(2) ${\it Originality}_i$ & 0.483 & 1 &  &  &  \\
(3) ${\it Significance}_i$ & 0.476 & 0.293 & 1 &  &  \\
(4) ${\it Correctness}_i$ & 0.564 & 0.053 & -0.077 & 1 &  \\
(5) ${\it Exposition}_i$ & 0.618 & 0.254 & 0.229 & 0.406 & 1 \\
\midrule
\multicolumn{6}{l}{\textit{Panel B: Accounting (41 papers)}} \\
(1) ${\it Overall}_i$ & 1 &  &  &  &  \\
(2) ${\it Originality}_i$ & 0.525 & 1 &  &  &  \\
(3) ${\it Significance}_i$ & 0.456 & 0.454 & 1 &  &  \\
(4) ${\it Correctness}_i$ & 0.534 & -0.130 & 0.007 & 1 &  \\
(5) ${\it Exposition}_i$ & 0.555 & 0.500 & 0.170 & 0.337 & 1 \\
\bottomrule
\end{tabular}
}

    \endgroup
    \label{tab:dim_corr}
\end{table}

\section{Variable Definitions}
\label{sec:appendix_variables}
{\footnotesize\renewcommand{\baselinestretch}{1}\selectfont
\begin{xltabular}{\linewidth}{lX}
    \caption{\textbf{Variable Definitions}\\This table defines the variables used in the analyses. Scores are the mean over four independent draws of \texttt{gpt-4o-2024-11-20}, or of \texttt{gpt-5-2025-08-07} in \autoref{tab:logit_gpt5}, at the default temperature; standardized versions have zero mean and unit variance across the papers of each regression. Publication outcomes are measured within two years of each paper's meeting.}
    \label{tab:var_des} \\
    \toprule
    Variable & Definition \\
    \midrule
    \endfirsthead
    \multicolumn{2}{l}{\textit{Table \thetable\ continued}} \\
    \toprule
    Variable & Definition \\
    \midrule
    \endhead
    \bottomrule
    \endlastfoot
    \multicolumn{2}{l}{\textit{Panel A: LLM referee scores}} \\
    ${\it Overall}_i$ & Overall score on a 0--100 scale, mean over draws. \\
    ${\it Originality}_i$ & Score on a 0--10 scale for whether the paper presents a genuinely new idea, angle, dataset, design, or synthesis relative to established academic knowledge \citep{newton2026}. \\
    ${\it Significance}_i$ & Score on a 0--10 scale for how important the question is and how large the contribution is if the results are correct. \\
    ${\it Correctness}_i$ & Score on a 0--10 scale for whether the analysis, empirical design, theoretical logic, and interpretation are scientifically convincing. \\
    ${\it Exposition}_i$ & Score on a 0--10 scale for how clear, well organized, and well communicated the paper is. \\
    \midrule
    \multicolumn{2}{l}{\textit{Panel B: Publication outcomes}} \\
    ${\it Top3Fin}_i$ & One if the paper is online at the \emph{Journal of Finance}, the \emph{Journal of Financial Economics}, or the \emph{Review of Financial Studies} within two years of the paper's meeting \citep{dai2023}. \\
    ${\it Top5Fin}_i$ & One if ${\it Top3Fin}_i = 1$ or the paper is online at the \emph{Journal of Financial and Quantitative Analysis} or the \emph{Review of Finance} \citep{dai2023}. \\
    ${\it Top10Fin}_i$ & One if ${\it Top5Fin}_i = 1$ or the paper is online at the \emph{Journal of Corporate Finance}, the \emph{Journal of Banking \& Finance}, \emph{Financial Management}, the \emph{Journal of Financial Markets}, or the \emph{Journal of Financial Intermediation} \citep{dai2023}. \\
    ${\it Top3Acc}_i$ & One if the paper is online at \emph{The Accounting Review}, the \emph{Journal of Accounting Research}, or the \emph{Journal of Accounting and Economics} within two years of the paper's meeting. \\
    \midrule
    \multicolumn{2}{l}{\textit{Panel C: Paper characteristics}} \\
    ${\it Log\ Length}_i$ & Natural logarithm of the number of tokens (o200k tokenizer) of the cleaned paper text. \\
    ${\it N\ Authors}_i$ & Number of authors listed on the conference program. \\
    ${\it NBER}_i$ & One if the paper was presented at a 2024 NBER finance meeting. \\
    \midrule
    \multicolumn{2}{l}{\textit{Panel D: Author reputation (Section~\ref{sec:appendix_reputation})}} \\
    ${\it Top20School}_i$ & One if any author's affiliation on the conference program is one of the twenty schools of the UT Dallas Top 100 Business School Research Rankings restricted to the \emph{Journal of Finance}, \emph{Journal of Financial Economics}, and \emph{Review of Financial Studies}, worldwide, 2019--2023 \citep{dai2023}; the same list is used for the accounting papers. \\
    ${\it Established}_i$ & One if any author of the paper has an article in the \emph{Journal of Finance}, \emph{Journal of Financial Economics}, or \emph{Review of Financial Studies} published before 2024 \citep{card2020}. \\
    ${\it Log\ Prior\ Top}_i$ & Natural logarithm of one plus the largest number of articles in the \emph{Journal of Finance}, \emph{Journal of Financial Economics}, or \emph{Review of Financial Studies} published before 2024 by any author of the paper, from OpenAlex; for the accounting papers, articles in \emph{The Accounting Review}, the \emph{Journal of Accounting Research}, or the \emph{Journal of Accounting and Economics}. \\
    ${\it Editorial}_i$ & One if any author of the paper is listed as an editor (editor, co-editor, senior or executive editor, editor-in-chief) or associate editor of the \emph{Journal of Finance}, \emph{Journal of Financial Economics}, or \emph{Review of Financial Studies} in any year from 2014 to 2024, the ten years up to the conference, from the journals' editorial-board pages and issue mastheads \citep{brogaard2014}; for the accounting papers, of \emph{The Accounting Review}, the \emph{Journal of Accounting Research}, or the \emph{Journal of Accounting and Economics}. \\
    ${\it Early\ Stage}_i$ & One if the first OpenAlex publication of every matched author of the paper is in 2014 or later, the ten years before the conference \citep[Table~11]{dai2023}. \\
    \midrule
    \multicolumn{2}{l}{\textit{Panel E: Look-ahead probe (Section~\ref{sec:appendix_lookahead})}} \\
    ${\it Claim}_i$ & One if, asked in a fresh session at temperature 0 for the publication status of the paper given its conference title, or its title and authors, the model answers that the paper is published or accepted; one variable per information set. \\
\end{xltabular}}

\end{document}